\documentclass[authoryear, twoside, a4paper, 11pt]{article}

\pdfoutput=1
\usepackage[english]{babel}
\usepackage[utf8x]{inputenc}
\usepackage[a4paper,top=3cm,bottom=2cm,left=3cm,right=3cm,marginparwidth=1.75cm]{geometry}
\usepackage{amsmath}
\usepackage{graphicx}
\usepackage{tabularx}
\usepackage{xcolor}
\usepackage{authblk}
\usepackage[round]{natbib}
\usepackage{subfig} 
\newcommand{\red}[1]{{\color{black} {#1}}}

\title{The radioscience LaRa instrument onboard ExoMars 2020 to investigate the 
rotation and interior of Mars}
\author[1,2]{V. Dehant}
\author[1]{S. Le Maistre }
\author[1]{R.-M. Baland }
\author[1]{N. Bergeot }
\author[1]{O. Karatekin }
\author[1]{M.-J. P\'eters }
\author[1]{A. Rivoldini }
\author[2,1]{L. Ruiz Lozano }
\author[1]{O. Temel }
\author[1]{T. Van Hoolst }
\author[1]{M. Yseboodt}
\author[1]{M. Mitrovic }
\author[3]{A.S. Kosov }
\author[4]{V. Valenta}
\author[5]{L. Thomassen}
\author[2]{S. Karki}
\author[2]{K. Al Khalifeh}
\author[2]{C. Craeye}
\author[6,7]{L.I. Gurvits }
\author[8]{J.-C. Marty }
\author[9]{S. Asmar}
\author[9]{W. Folkner }
\author{and the LaRa team (http://lara.oma.be/team)}

\affil[1]{Royal Observatory of Belgium, Brussels, 3 avenue Circulaire, B1180 
Brussels, Belgium, v.dehant@oma.be}
\affil[2]{Universit\'e catholique de Louvain, Belgium}
\affil[3]{IKI - Space Research Institute of Russian Academy of Sciences, Moscow, 
Russia}
\affil[4]{European Space Research and Technology Centre (ESTEC), ESA, The 
Netherlands}
\affil[5]{AntwerpSpace, OHB Company, Belgium}
\affil[6]{JIVE - Joint Institute for VLBI ERIC (European Research Infrastructure 
Consortium), The Netherlands}
\affil[7]{Department of Astrodynamics and Space Missions, Delft University of 
Technology, The Netherlands}
\affil[8]{Observatoire Midi-Pyr\'en\'ees, GRGS, CNES, France}
\affil[9]{Jet Propulsion Laboratory, California Institute of Technology, 
Pasadena, USA}

\date{Accepted in PSS, October 2019}

\begin{document}
\maketitle

\begin{abstract}
LaRa (Lander Radioscience) is an experiment on the ExoMars 2020 mission that 
uses the Doppler shift on the radio link \red{due to the motion of} the ExoMars 
platform \red{tied to the surface of Mars with respect to} the Earth \red{ground 
stations (e.g.~the deep space network stations of NASA)}, in order to precisely 
measure the relative velocity of the lander on Mars with respect to the Earth. 
The LaRa measurements shall improve the understanding of the structure and 
processes in the deep interior of Mars by obtaining the rotation and orientation 
of Mars with a better precision compared to the previous missions. In this 
paper, we provide the analysis done until now for the best realization of these 
objectives. \red{We explain the geophysical observation that will be reached 
with LaRa (Length-of-day variations, precession, nutation, and possibly polar 
motion).
We develop the experiment set up, which includes the ground stations on Earth 
(so-called ground segment). We describe the instrument, i.e.~the transponder and 
its three antennas. We further detail the link budget and the expected noise 
level that will be reached. 
Finally, we detail the expected results, which encompasses the explanation of 
how we shall determine Mars' orientation parameters, and the way we shall deduce 
Mars' interior structure and Mars' atmosphere from them.
Lastly, we explain briefly how we will be able to determine the Surface platform position.}
\end{abstract}

\section{Introduction}
Shortly after their formation, Earth and Mars might have been very similar. 
Nowadays, those neighboring planets show some differences proving that they have 
evolved differently. For example, Mars has a tenuous atmosphere mainly made of 
CO$_2$ and containing \red{in average} almost no oxygen. An important part of 
its surface has been generated a long time ago \red{(before 4 Ga)} and shows no 
sign of regional-scale recent alteration, while the Earth’s surface is 
continuously recycled through plate tectonics. Mars is monoplate. Mars is 
presently and since the Noachian ($\approx$3.5 Ga), a very dry planet. All this 
indicates that both planets, while being quite similar terrestrial planets, also 
differ internally. For instance, although it is known \red{from observation of 
the tidal $k_2$ Love number from orbiters (for the first time determined by 
\citet{Yoder:2003fk})} that Mars’ iron-rich core has a radius of about 1794 ± 
65~km at one sigma (range [1600 km, 1990 km] for 3 sigma, 
\citep{Smrekar:2019aa}), its composition and the planet’s thermal state are not 
well known. The present uncertainty on the core radius and the not-well-known 
lower mantle temperature have major consequences for the understanding of the 
global interior structure and dynamics and of the planetary evolution. For 
example, for a bridgmanite lower mantle to exist in Mars, the \red{liquid} core 
must be sufficiently small. If present, the endothermic ringwoodite-bridgmanite 
phase transition would have a focusing effect on mantle convection, which could 
help formation and maintaining Tharsis 
\citep{Harder:1996aa,van-Thienen:2005aa,van-Thienen:2006aa}. The thermal state 
and composition of the core are also important for the history of the magnetic 
dynamo, which in turn could have important consequences for the retention of the 
atmosphere and the possible habitability of the surface early in Mars’ history.
\\

\begin{figure}[!ht]
\centering
\includegraphics[width=0.7\textwidth]{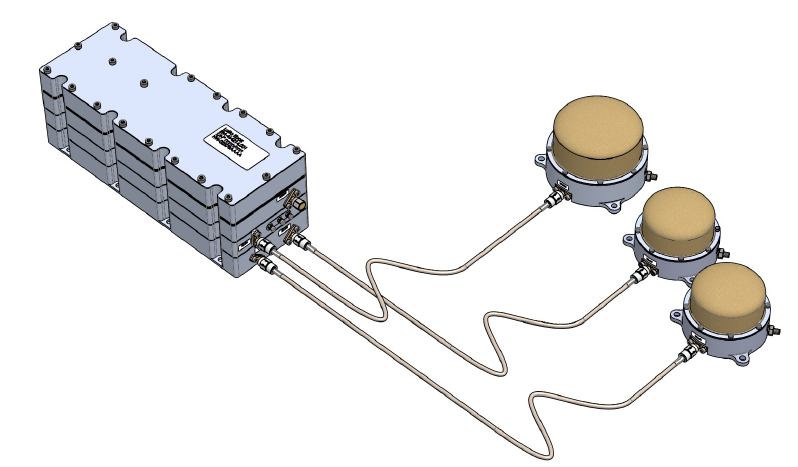}
\caption{LaRa assembly consisting of a transponder and 3 antennas that are 
interconnected via 3 coaxial cables.}
\label{figharness}
\end{figure}

We are interested in investigating how and why Mars differs from Earth. To that 
aim, we have designed an experiment addressing the deep interior, essentially 
the core, and the atmosphere dynamics: the LaRa (Lander Radioscience) 
experiment. LaRa is the generic name for the transponder and its antennas (see 
Fig.~\ref{figharness}) that will be included in the payload elements on the 
Surface Platform (SP, \red{also called Kazachok Platform}) of the ExoMars 2020 
mission led by ESA and Roscosmos. The experiment uses a coherent X-band 
transponder on Mars to obtain two-way Doppler measurements, i.e.~to measure the 
line-of-sight velocity variations between the SP and the Earth \red{station}, 
and is designed to obtain measurements over at least one Martian year \red{(687 
Earth days)}. These Doppler measurements, in conjunction with other previous or 
simultaneous direct-to-Earth (DTE) radio link measurements, will be used to 
obtain Mars’ rotational behavior (precession, nutations, length-of-day (LOD) 
variations, and polar motion). More specifically, measuring the relative 
position of the SP on the surface of Mars with respect to the terrestrial ground 
stations allows reconstructing Mars’ time-varying orientation and rotation in 
\red{inertial} space, knowing the Earth's orientation precisely 
\red{(i.e.,~Earth's precession, nutation, polar motion, and length-of-day 
variations)}. These precise measurements are used to determine the inertia of 
the whole planet (mantle plus core), the moment of inertia (MoI) of the core, 
and the global-scale seasonal \red{changes} of the atmosphere (variation of the 
angular momentum and inertia due to the seasonal mass transfer of CO$_2$ between 
the atmosphere and ice caps, \red{thus informing about the atmosphere 
dynamics}).\\
\\
The LaRa experiment will be conducted jointly with the other experiments of the 
ExoMars missions as well as of the NASA InSight mission (Interior Exploration 
using Seismic Investigations, Geodesy and Heat Transport) in order to obtain the 
maximum amount of information about the interior of Mars and consequently on its 
formation and evolution, in accordance with the ExoMars objective to investigate 
the planet’s deep interior to better understand Mars’ evolution and 
habitability, as well as to investigate the Martian atmosphere.\\
\\
The paper is organized as follows: Section~\ref{sectionscirationale} provides 
the scientific rationale, including Mars' rotation model, interior structure, 
and atmosphere dynamics. 
Section~\ref{sectionLaRaexp} describes the history of the LaRa experiment, its 
objectives, the LaRa team, the transponder and the antennas, the ground segment, 
the link budget, the measurements, and the operations. The experiment involves 
an important ground segment, which includes the Deep Space Network (DSN) and the 
ESA TRACKing stations (ESTRACK). Complementary observation will be performed by 
VLBI (Very Long Baseline Interferometry) radiotelescopes in the frame of the 
Planetary Radio Interferometry and Doppler Experiment (PRIDE) (see 
Section~\ref{pride}).
Section~\ref{sectionMOP} summarizes the expected results of the LaRa experiment: 
the improvement expected for the MOP (Mars Orientation and rotation Parameters) 
determination and the consequences for our understanding of Mars' interior 
structure and atmosphere dynamics. The capability of LaRa to accurately locate 
the lander at the surface of Mars after few days at the surface is also 
addressed in this section.

\section{Mars rotation}
\label{sectionscirationale}

The variations in the rotation of Mars can be separated into (1) the variations 
in the rotation about the spin axis, (2) the variations of spin axis with 
respect to inertial space 
(variations in orientation of the spin axis in space), and (3) the changes of 
the spin axis relative to body-fixed axes. LaRa will improve current estimates 
of those variations or measure them for the first time, to yield information on 
the interior structure and atmosphere of Mars, as described in the introduction.

\begin{table}[!htb]
\caption{List of acronyms used in the paper.}
\centering
\begin{tiny}
$\begin{array}{|l|l|}
\hline
\text{ADEV} & \text{Allan deviation} \\ 
\text{AM} & \text{Angular Momentum} \\
\text{ATDF} & \text{Ascii Test Data Format} \\
\text{BELSPO} & \text{Belgian Federal Science Policy Office} \\
\text{BF} & \text{Body Frame} \\
\text{BIP} & \text{Acronym for the Onboard Information and Memory unit in 
Russian} \\
\text{CDR} & \text{Critical Design Review} \\
\text{CW} & \text{Chandler wobble} \\
\text{DSN} & \text{Deep Space Network} \\
\text{DTE} & \text{Direct-to-Earth} \\
\text{DW} & \text{Mantle mineralogy composition from Taylor et al. (2013) 
starting from the } \\ 
& \text{model of Dreibus \& W\"anke (1984, 1987) and W\"anke \&  Dreibus (1988, 
1994), } \\ 
& \text{and adding further laboratory data, see \citep{Taylor:2013fk} and 
references in that paper} \\
\text{ESOC} & \text{European Space Operations Centre} \\
\text{ESTRACK} & \text{ESA's European Space TRACKing network} \\
\text{EH45} & \text{Mantle mineralogy composition from Sanloup et al. (1999) 
using intermediate} \\ 
& \text{chondrites between H ordinary chondrites and EH enstatite chondrites 
\citep{Sanloup:1999gj}} \\ 
& \text{(EH45:E55)} \\
\text{EVN} & \text{European VLBI Network} \\
\text{FCN} & \text{Free Core Nutation} \\
\text{FICN} & \text{Free Inner Core Nutation} \\
\text{GEP} & \text{Geophysics and Environmental Package} \\
\text{GINS} & \text{G\'eod\'esie par Int\'egrations Num\'eriques Simultan\'ees} 
\\
\text{ICRF} & \text{International Celestial Reference Frame} \\
\text{IF} & \text{Inertial Frame} \\
\text{IKI RAS} & \text{Space Research Institute of the Russian Academy of 
Sciences} \\
\text{InSight} & \text{Interior Exploration using Seismic Investigations, 
Geodesy and Heat Transport} \\
\text{ITU} & \text{International Telecommunication Union} \\
\text{JIVE} & \text{Joint Institute for VLBI European Research Infrastructure 
Consortium} \\
\text{JUICE} & \text{JUpiter ICy moons Explorer} \\
\text{LaRa} & \text{Lander Radioscience} \\
\text{LF} & \text{Mantle mineralogy composition from Lodders and Fegley (1997) 
\citep{ Lodders:1997rv}} \\
\text{LO} & \text{Local oscillator} \\
\text{LOD} & \text{Length of Day} \\
\text{MA} & \text{Mantle mineralogy composition from Morgan and Anders (1979) 
\citep{ Morgan:1979mm}} \\
\text{MM} & \text{Mantle mineralogy composition from Mohapatra and Murty (2003) 
\citep{ Mohapatra:2003zc}} \\
\text{MoI} & \text{Moment of inertia} \\
\text{MOP} & \text{Mars Orientation and rotation Parameters} \\
\text{NEIGE} & \text{NEtlander Ionosphere and Geodesy Experiment} \\
\text{ODF} & \text{Operation Data File - Orbit Data File } \\ 
\text{PM} & \text{Polar Motion} \\
\text{PRIDE} & \text{Planetary Radio Interferometry and Doppler Experiment} \\
\text{PRODEX} & \text{PROgramme for the Development of scientific EXperiments} 
\\
\text{RGS} & \text{Russian ground station} \\
\text{RHU} & \text{Radioisotope Heater Unit} \\
\text{RISE} & \text{Rotation and Interior Structure Experiment} \\
\text{ROB} & \text{Royal Observatory of Belgium} \\
\text{SEIS} & \text{The ExoMars seismometer} \\
\text{SNR} & \text{Signal-to-noise ratio} \\
\text{SP} & \text{Surface Platform} \\
\text{SSPA} & \text{Solid-State Power Amplifier} \\
\text{TDM} & \text{Technical Data Measurement file - Tracking Data Message} \\
\text{TNF} & \text{Tracking and Navigation data File} \\ 
\text{TU Delft} & \text{Delft University of Technology} \\
\text{TVAC} & \text{Thermal and Vacuum test} \\
\text{VLBI} & \text{Very Long Baseline Interferometry} \\
\hline
\end{array}$
\end{tiny}
\label{tablacro}
\end{table}

\subsection{Rotation model of Mars}
\begin{figure}[!ht]
\centering
\includegraphics[width=0.7\textwidth]{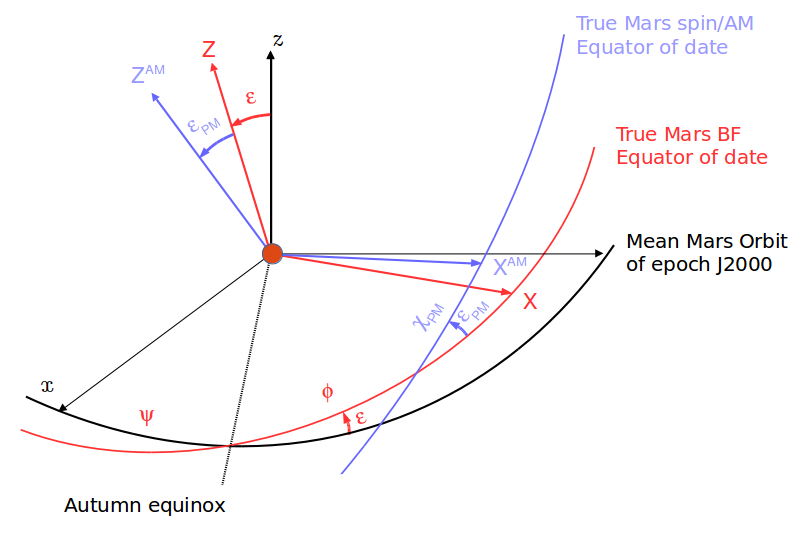}
\caption{Euler angles (defined as prograde angles) between the Martian rotating 
BF (axes $XYZ$) and the IF associated to the Martian mean orbit at epoch J2000.0 
(axes $xyz$). The $X$-axis of the BF is chosen as the prime meridian defined in 
the IAU (International Astronomical Union) convention \citep{Archinal:2018aa}. 
$\psi$ is measured from the $x$-axis to the autumn equinox, $\phi$ is measured 
from the equinox to the $X$-axis, and $\varepsilon$ is the angle from the 
$z$-axis to the $Z$-axis, or the inclination of the BF equator over the IF 
equator. The spin axis does not coincide with the figure axis of Mars. They are 
inclined by $\varepsilon_{PM}$ to each other, and $\chi_{PM}$ is measured from 
the node of the spin equator over the BF equator to the $X^{AM}$-axis (the 
subscript $PM$ stands for \textit{Polar Motion}). The angular momentum axis can 
be considered as aligned to the spin axis \citep{Bouquillon:1999BS}.}
\label{eulerangles}
\end{figure}
The rotation of Mars is described by the temporal evolution of the Euler angles 
from an Inertial Frame (IF) to the rotating Body Frame (BF) of Mars (i.e.,~a 
frame tied to the planet), or from the IF to a reference frame attached to the 
angular momentum (AM) axis, which is very close to the spin axis 
\citep{Bouquillon:1999BS}. Those angles are the node longitude $\psi$, the 
obliquity $\varepsilon$, and the rotation angle $\phi$ (see 
Fig.~\ref{eulerangles}). The IF is here chosen to be associated with the mean 
orbital plane of Mars at a reference epoch (J2000.0). The BF is attached to the 
principal axes of inertia of the planet, and therefore associated with the 
equator of figure of Mars. In models of the rotation of Mars, the planet is 
often first considered to respond rigidly to rotational forcing. Small non-rigid 
contributions are added afterwards. For a non-rigid Mars, the BF is attached to 
the mean principal axes of the mantle. In the next three subsections, the 
different rotation characteristics of Mars will be defined. The relation with 
the interior structure and atmosphere is described in sections \ref{RotandInt} 
and \ref{RotandAtm}.

\subsubsection{Length-of-day variations }
The temporal variations of the rotation angle $\phi$ can be decomposed into a 
uniform rotation and periodic variations:
\begin{equation}
 \phi=\phi_0+\Omega_0 t+\Delta\phi
\end{equation}
where \red{t is the time past the J2000 epoch,} $\phi_0$ is such that the 
$X$-axis points in the direction of the prime meridian at J2000.0, and 
$\Omega_0=350^{\circ}.89/$day is the constant rate of uniform rotation 
\citep{Archinal:2018aa}. $\Delta\phi$ are the variations induced mainly by the 
exchanges of angular momentum between the atmosphere and the planet due to the 
seasonal CO$_2$ sublimation and condensation processes at polar caps, mass 
re-distributions in the atmosphere, and seasonally changing winds. $\Delta\phi$ 
can be written as a series
\begin{equation}
\label{DeltaPhi} \Delta\phi=\sum_m \phi_{m}^c \cos(f_m t)+\phi_{m}^s \sin(f_m 
t),
\end{equation}
with $f_m$, the main frequencies associated to seasonal processes ({\it e.g.} 
annual, semi-annual), and $\phi_{m}^{c/s}$, the amplitudes associated with each 
frequency $f_m$ (see Section \ref{RotandAtm}). 

Since deviations from the orientation given by the uniform rotation imply 
changes in the length of a day, it is customary to use length-of-day (LOD) 
variations $\Delta LOD=2\pi/(\frac{d \phi}{dt})-\frac{2\pi}{\Omega_0}$.  The 
mean LOD variation, averaged over many years  has been measured by combining 
Viking and Pathfinder data with the tracking of Mars orbiters: Mars Global 
Surveyor (MGS), Mars Odyssey (ODY) and Mars Reconnaissance Orbiter (MRO) 
\citep{Konopliv:2006pi, Konopliv:2011fk, Konopliv:2016aa}.  LaRa will improve 
current estimates of the LOD variations (known to within about 15\%, 
\citet{Konopliv:2011fk}), give insight on interannual variations and provide 
global constraints on the distribution of atmospheric mass, angular momentum, 
and the ice caps.

\subsubsection{Precession and nutations}
Since Mars is tilted in space (the $Z$-axis of the BF is not parallel to the 
$z$-axis of the IF ($\varepsilon\simeq 25^\circ$)), flattened and rotating, it 
reacts as a spinning top to the gravitational torque exerted by the Sun. Because 
of the orbital eccentricity of Mars and of the orbital changes due to 
interactions with other planets, the Solar gravitational torque on Mars changes 
periodically with time. The other planets, as well as Phobos and Deimos, the two 
natural satellites of Mars, also exert direct gravitational torques on Mars.

\begin{figure}[!ht]
\centering
\includegraphics[width=0.49\textwidth]{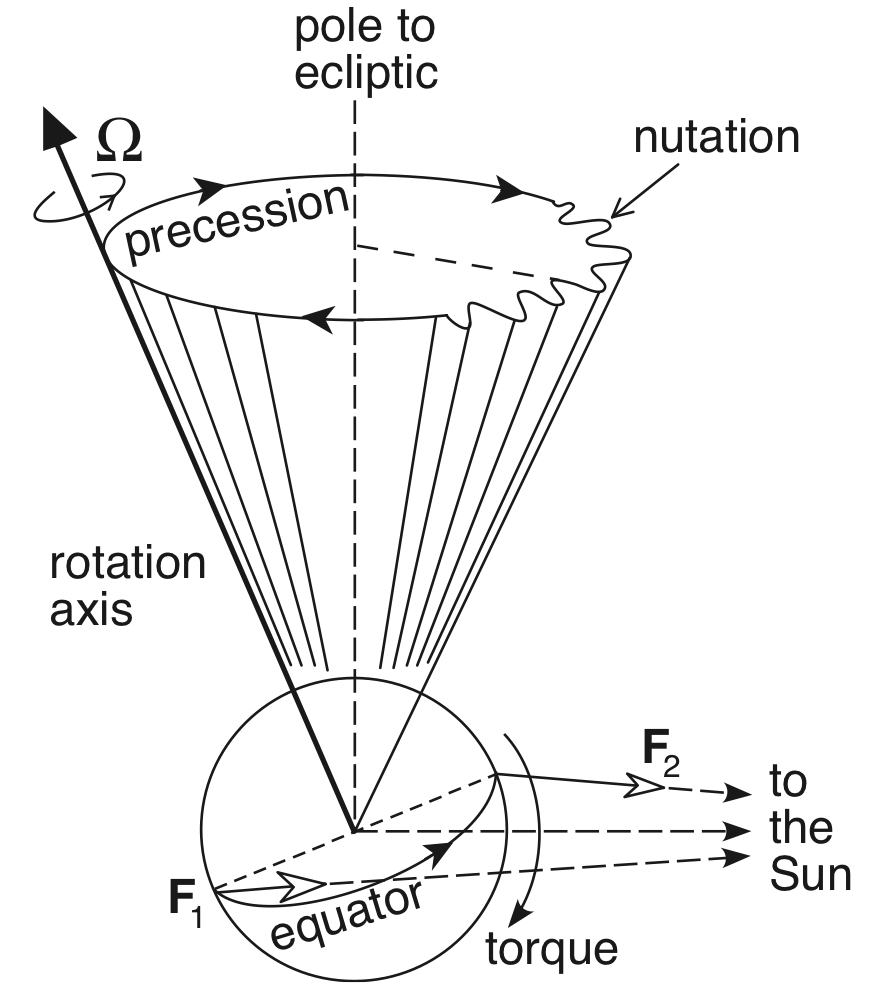} 
\caption{Precession and nutations of the rotation axis about the orbit pole 
\citep{Lowrie:2011zr}}
\label{figprecnut}
\end{figure}

As a consequence, the angles $\psi$ and $\varepsilon$, change with time at 
various time-scales. The angle $\psi$ can be decomposed into a slow uniform 
precession around the $z$-axis (see Fig.~\ref{figprecnut}) at rate $\dot \psi$ 
and periodic nutations in longitude $\Delta\psi$. Referring the nutations with 
respect to the J2000 mean equinox, the longitude is defined as $\psi=\dot \psi t 
+ \Delta\psi$. The angle $\varepsilon$ is the sum of the J2000.0 value 
$\varepsilon_0$ and of periodic nutations in obliquity $\Delta \varepsilon$. The 
time needed to perform one precession cycle around the orbit normal is about 
171,000 years ($\dot\psi\simeq -7.6$ arcsecond/year). A first objective of LaRa 
is to very accurately determine the precession rate. Since $\dot\psi$ is 
inversely proportional to the polar principal moment of inertia (MoI), LaRa will 
be able to provide accurate constraints on the interior structure. This has 
already been performed by using the Viking Lander radio link 
(\cite{Yoder:1997jo}) together with Pathfinder spacecraft radio-link 
(\cite{Folkner:1997mb}), combined with orbiter data (\cite{Konopliv:2006pi, 
Konopliv:2011fk, Konopliv:2016aa}), as well as by using the MERs (Mars 
Exploration Rovers) when fixed (\cite{Kuchynka:2013uq, Le-Maistre:2013aa}). The 
present uncertainty on the precession rate (\cite{Konopliv:2016aa}) of 
2.1~mas/year, corresponding to a period change of about 47 years, will be 
improved by an order of magnitude (up to 0.3~mas/year) by using LaRa and RISE 
(Rotation and Interior Structure Experiment, onboard the InSight NASA mission 
landed on Mars in 2018) data, see Table \ref{tab:rotmod}.

The resulting motion due to precession and nutation is wiggly as illustrated in 
Fig.~\ref{figprecnut}. In practice, the variables $\psi$ and $\varepsilon$ are 
often related to $(\delta x,\delta y)$, the projection of the trajectory of a 
unit vector along the figure axis onto the J2000 mean BF equator 
\citep{Defraigne:1995kh}. The mean BF is defined so that it follows the uniform 
precession but not the nutations. At first order in small variations, around 
J2000, that projection can be written as 
\begin{eqnarray}
 (\delta x,\delta y)&=&(\sin\varepsilon_o (\dot \psi t+\Delta \psi), -\Delta 
\varepsilon).
\end{eqnarray}
where the precession rate $\dot\psi$ is usually omitted to focus on the 
nutations. It can be expressed as series of prograde and retrograde circular 
motions
\begin{equation}
\left\lbrace \begin{array}{c}
 \delta x\\
 \delta y
 \end{array}\right\rbrace=\sum_j \left(\mathcal{P}_j \left\lbrace 
\begin{array}{c}\cos\\\sin\end{array}\right\rbrace 
(f_j\,t+\pi_{j}^{0})+\mathcal{R}_j \left\lbrace 
\begin{array}{c}\cos\\\sin\end{array}\right\rbrace(-f_j\,t-\rho_{j}^{0})\right)
\end{equation}
whose amplitudes $\mathcal{P}_j$ and $\mathcal{R}_j$ and phases $\pi_{j}^{0}$ 
and $\rho_{j}^{0}$ at J2000 are related to the amplitudes and phases of similar 
series for the longitude and obliquity nutations such as those computed by 
\citep{Reasenberg:1979fk} for Mars figure axis, or those of 
\citet{Bouquillon:1999BS} or \citet{Roosbeek:1999lh} for Mars AM axis, and 
expressed as
\begin{equation}
\left\lbrace \begin{array}{c}
 \Delta \psi\\
 \Delta \varepsilon
 \end{array}\right\rbrace=\sum_j \left( \left\lbrace 
\begin{array}{c}\psi_j^{c}\\\varepsilon_j^{c}\end{array}\right\rbrace \cos 
\varphi_j+ \left\lbrace 
\begin{array}{c}\psi_j^{s}\\\varepsilon_j^{s}\end{array}\right\rbrace 
\sin\varphi_j\right)
\end{equation}
The arguments $\varphi_j=f_j t+\varphi_{j}^{0}$ are determined by the orbital 
motion of the perturbing bodies and by the rotation and orbit of Mars and can be 
expressed as linear combinations of fundamental arguments, such as the mean 
longitudes of Earth, Mars, Jupiter, and Saturn, and the node of Phobos and 
Deimos. The main terms are related to arguments which are multiples of the mean 
longitude of Mars, the largest nutation being semi-annual with an amplitude of 
about $500$~mas for the prograde component \citep{Baland:2019in}. The main 
nutation terms have first been measured by \cite{Borderies:1980kc} with large 
uncertainties (at 50-100\% level). \cite{Le-Maistre:2013aa} and 
\cite{Le-Maistre:2018aa} recently published new estimates of those terms with an 
accuracy of few tens of mas, still insufficient to constrain the core 
characteristics. The  radioscience experiments (RISE and LaRa) will further 
improve the main nutation terms of Mars.

\subsubsection{Polar motion }
Polar motion describes differential motion between the figure and spin/AM axes. 
The projection of the spin axis in the BF is often denoted $(X_P,-Y_P)$ (the 
minus sign comes from the convention \red{used for Earth polar motion}) and can 
be expressed in terms of two angles as $(-\varepsilon_{PM}\sin 
\chi_{PM},-\varepsilon_{PM}\cos\chi_{PM})$, with $\chi_{PM}$ and 
$\varepsilon_{PM}$, the angles from the BF to the AM frame (see 
Fig.~\ref{eulerangles}). Forced polar motion is due to the redistribution and 
motion of mass elements at seasonal time scales. The existence of free 
oscillations, like the Chandler Wobble (CW), if excited, induce an additional 
component to be added to the seasonal components. Forced polar motion $X_P$ and 
$Y_P$ can therefore be expressed as series of seasonal periodic components, 
similarly as the variations in rotation angle of Eq.~(\ref{DeltaPhi}), with 
amplitudes expected to be in the range between $0$ and $15$~mas 
\citep{Defraigne:2000ss, Van-den-Acker:2002qq} (see section \ref{RotandAtm}). 
The atmosphere could also excite the CW, a normal mode for the motion of the 
spin axis with respect to the figure axis. The period of the Martian CW is 
expected to be around 200 days for the non-hydrostatic case and 220 days for the 
hydrostatic case \citep{Van-Hoolst:2000if, Van-Hoolst:2000kb}, and its amplitude 
could be of the order of 10 to 100~mas \citep{Dehant:2006aa}. This amplitude can 
vary significantly from one year to another as has been observed on the Earth 
\citep{Seitz2005JGRB}. LaRa, which will be at a higher latitude than RISE, is 
better suited to detect Polar Motion and/or the CW \citep{Yseboodt:2017if, 
Yseboodt:2017ab}. However, LaRa latitude is still not very high and the CW 
detection will depend on its actual amplitude.

\subsubsection{The rotation matrix}
The rotation matrix from the BF to the IF can be expressed in terms of the Euler 
angles ($\psi$, $\varepsilon$, $\phi$) as \citep{Reasenberg:1979fk}
\begin{equation}
 \vec M=R_z(-\psi).R_x(-\varepsilon).R_z(-\phi).
\end{equation}
More conveniently, the transformation can be performed by using the three types 
of rotational variations as \citep{Folkner:1997dw,  Konopliv:2006pi, 
Le-Maistre:2012ff}
\begin{equation}
 \vec M=R_z(-\psi) R_x(-\varepsilon) R_z(-\phi) R_y(X_P) R_x(Y_P)
\end{equation}
where $(\psi,\varepsilon,\phi)$ are the AM/spin Euler angles.

\subsection{Rotation and Interior structure}
\label{RotandInt}
The time-dependent tidal forcing exerted by the Sun, the other planets, and 
Phobos and Deimos on the flattened Mars induces periodic nutations of its 
rotation axis \citep[e.g.][]{Dehant:2015bs}. Nutations can be resonantly 
amplified when their frequency  is close to that of a rotational normal mode, in 
particular the CW, the Free Core Nutation (FCN), and the Free Inner Core 
Nutation (FICN).  The CW is a motion of the rotation axis around the figure axis 
and occurs when both axes have an offset from each other. The FCN mode describes 
a relative rotation of the liquid core with respect to the mantle. The FICN is a 
relative motion of the rotation axis of a solid inner core with respect to those 
of the mantle and liquid core and therefore can only affect nutation if Mars has 
a solid inner core. The period of the three normal modes and the amplification 
they induce depend on the moments of inertia of the planet, core, and inner core 
and on their capacity to deform in response to rotation rate variations and 
tidal gravitational forcing \red{(elastic deformations due to external tidal 
potential and rotational centrifugal potential)} 
\citep[e.g.][]{Mathews:1991pi,Dehant:2015bs}.\\

The CW has a long period in a rotating frame tied to the planet, while the FCN 
and FICN, as the nutations, have nearly diurnal periods in that frame and long 
periods in the inertial frame. Therefore, at the first order, the influence of 
the CW on nutations is very small and usually neglected.\\

Although the resonant amplification induced by the FICN could in principle be 
detected by LaRa \citep{Defraigne:2003gr}, 
constraints about the core composition deduced from geodesy data 
\citep{Rivoldini:2011fk,Khan:2018aa} would require unrealistic low core 
temperatures \citep{Fei:2000ei,Plesa:2016aa} (close to the Fe-S eutectic 
temperature) for inner-core formation to occur. For this reason, we do not 
consider Mars models with an inner core.\\

For a tri-axial planet with a liquid core, to first order the frequency of the 
CW and FCN in a frame co-rotating with Mars can be written as 
\citep{Chen:2010kx}
\begin{align}
\omega_\mathrm{CW}&=\Omega \frac{A}{A_m} \sqrt{(\alpha-\tilde 
{\kappa})^2-\frac{1}{4}\beta^2} \label{eq:CW}\\
\omega_\mathrm{FCN}&=-\Omega \frac{A}{A_m} \sqrt{(\alpha_f-\tilde 
{\beta})^2-\frac{1}{4}\beta_f^2}-\Omega \label{eq:FCN}
\end{align} 
where $\Omega$ is the rotation frequency of Mars, $A$ the equatorial moment of 
inertia, $A_m$ the equatorial moment of inertia of the mantle, $\alpha$ and 
$\beta$ are the dynamic polar and equatorial flattenings of the planet and the 
corresponding symbols indexed with $_f$ are those of the core. The compliance 
$\tilde{\kappa}$  characterizes the yielding of the planet to tidal forcing and 
the compliance $\tilde{\beta}$ quantifies the core's capacity to deform due to 
rotation rate variations \citep[e.g.][]{Dehant:2015bs}. The dynamical polar and 
equatorial flattening are:
\begin{equation}
    \alpha=\frac{C-\bar{A}}{\bar{A}}\quad,\quad \beta=\frac{B-A}{A},
\end{equation}
where $A$, $B$, and $C$ are the principal moments of inertia of the planet and 
$\bar{A}=\frac{1}{2}(A+B)$. The dynamical flattening of the core are defined in 
a similar fashion.

If the planet's core equatorial flattening is small compared to its polar 
flattening, which is likely the case for Mars (Wieczorek et al 2019), then the 
frequency of the FCN can be calculated from the equation for the bi-axial case 
\citep{Van-Hoolst:2002aa} that is correct up to first order in flattening of the 
core. Corrections due to triaxiality are below 1~day in the inertial frame 
\citep{Van-Hoolst:2002aa} while the range of possible periods for the FCN is 
[$-$230,$-$280] days.  The FCN frequency for a bi-axial case writes
\begin{equation}
\omega_\mathrm{FCN}=-\Omega \frac{A}{A_m} (\alpha_f- 
\tilde{\beta})-\Omega. \label{eq:FCN2axial}
\end{equation} 
\\
The wobble of Mars can be calculated from the Liouville equations for a 
deformable planet with a liquid core \citep[e.g.][]{Dehant:2015bs}. By dividing 
this wobble with that of a rigid Mars, a \emph{transfer function} ($T_F$) 
depending only on the interior structure of the planet is obtained. From the 
transfer function, the nutation of Mars can be calculated from the rigid 
nutation that only depends on the planet's principal moments of inertia and on 
the external tidal forcing.

The prograde and retrograde nutation amplitudes $\mathcal{P}'$ and 
$\mathcal{R}'$ at a given frequency read as
\begin{equation}
\mathcal{P}'(\omega)=T_F(\omega)\mathcal{P}(\omega),\quad 
\mathcal{R}'(\omega)=T_F(-\omega)\mathcal{R}(\omega),
\end{equation}
where $\mathcal{P}$ and $\mathcal{R}$ are the rigid prograde and retrograde 
nutations. The amplitudes and frequencies of the principal rigid nutations are 
given in Tab.~\ref{tab:rotmod}. The transfer function in the frequency band of 
the rigid nutations can be calculated from
\begin{equation}
	T_F(\omega)=1+\frac{\bar{A}_f}{\bar{A}_m} 
\left(1-\frac{\tilde{\gamma}}{\alpha}  \right) 
\frac{\omega}{\omega-\omega_\mathrm{FCN}}=1+F 
\frac{\omega}{\omega-\omega_\mathrm{FCN}}, \label{eq:TF}
\end{equation} 
where $\bar{A}_m$ and $\bar{A}_f$ are the average equatorial moments of inertia 
of the mantle and core, $\omega$ and $\omega_{FCN}$ are frequencies with respect 
to the inertial frame, $\tilde{\gamma}$ describes the response of the core to 
tidal forcing, and $F$ is the liquid core amplification factor. For a biaxial 
planet Eq.~(\ref{eq:TF}) is correct to first order in flattening and since this 
expression is not directly dependent of the dynamic flattening of the core it 
also provides a precise approximation for the transfer function of a triaxial 
planet.

\subsection{Rotation and atmosphere dynamics} 
\label{RotandAtm}

The exchange of angular momentum between the fluid layers and the solid planet 
is the main cause for the variations of the rotation of terrestrial planets in 
seasonal time scales \citep{Karatekin:2011uq}. The atmospheric angular momentum 
 variations are directly linked with the three climate cycles of Mars:  CO$_2$, 
dust, and water cycles. The most important part of those variations is due to 
the  CO$_2$ cycle. The atmosphere of Mars is mainly controlled by the seasonal 
changes in the polar icecaps, resulting from the sublimation and condensation 
process of  CO$_2$. During the winter hemisphere, the temperature reaches the 
frost point temperature of  CO$_2$, which condensates and creates  CO$_2$ 
deposits on the surface. In return, during the summer in the same hemisphere, 
the  CO$_2$ polar cap sublimates back into the atmosphere. Both  H$_2$O ice and 
 CO$_2$ ice are observed unambiguously in visible and near Infrared wavelength 
range \citep{Langevin2006Natur}.  Contamination of  surface ices  by dust can 
make  the observations of polar caps evolution more challenging in some seasons 
\citep{Langevin2007JGRE}. The  CO$_2$ and H$_2$O sublimation and condensation 
processes are related with the airborne dust. The dust in the atmosphere of 
Mars, alters the radiative heat transfer in the atmosphere, which then strongly 
affects the atmospheric circulation.  The amount of dust on the surface also 
changes the surface albedo and thermal inertia, leading to different seasonal 
and diurnal surface temperature variations \citep{Kahre2010Icar}. In addition 
to dust, water ice clouds play an important role on the radiative balance in 
Mars’ atmosphere despite being a minor component of the atmosphere. Water ice 
clouds contribute to radiative heat transfer mainly in the infrared band, 
affecting the surface heat balance indirectly. It is also found that during 
northern hemisphere summer, the presence of water ice and dust clouds enhance 
the polar warming \citep{Kahre2015Icar}. Therefore, in the frame of the present 
study and in view of the high level of accuracy that we will obtain the LaRa 
data, it is not appropriate to investigate 
the CO$_2$ cycle decoupled from the water and dust cycles, which can cause 
temporary variations in the CO$_2$ cycle. 
Furthermore, the seasonal and interannual variations in these cycles directly 
affects the zonal mean transport, or the atmospheric angular momentum in other 
words.

Growth and retreat of the North seasonal cap is shown to be repeatable within 
1–2 degrees equivalent latitude, whereas the South seasonal cap  presents 
noticeable variability  in non-dusty years \citep{Piqueux2015Icar}. Dust storms 
may have  significant impact on the recession and growth rates of both  polar 
caps.  Atmospheric temperatures and dust loading are the primary source of 
variability in an otherwise remarkably repeatable cycle of seasonal cap growth.

Seasonal angular momentum and rotation variations of Mars estimated using the 
assimilated observations (i.e.,~reanalysis of fundamental atmospheric and 
surface variables) over Martian years  25 and 26 indicated that the effect of 
dust on Mars rotation could be visible in Mars year 25 when a global dust storm 
occurred \citep{Karatekin:2014aa}.  The dust storm in Mars year 28 had a more 
significant affect on the recession and extent of  caps compared to the previous 
dust storm  \citep{Piqueux2015Icar}. Variations in the rotation of Mars \red{are 
directly related to} seasonal and interannual variability of the ice caps and 
\red{their observation are thus} important for the understanding of Mars’ 
current atmospheric dynamics \citep{Karatekin:2005aa}.

The angular momentum exchange between the surface and the atmosphere, alters the 
planetary rotation, causing variations up to 0.4 millisecond (or equivalently 
10~m on the equator or 620~mas in terms of angular amplitude) 
in Martian LOD over seasonal time scales whereas the polar motion effect is 
predicted to be in the order of tens of milliarcsecond \citep{Karatekin:2011uq}. 
The predicted  amplitudes of seasonal LOD variations from  general circulation 
models with different prescribed dust scenarios as well as with assimilated 
observations are within the current observation uncertainties 
\citep{Konopliv:2011uq}.

\section{The LaRa experiment}
\label{sectionLaRaexp}
\begin{figure}[!ht]
\centering
\includegraphics[width=0.7\textwidth]{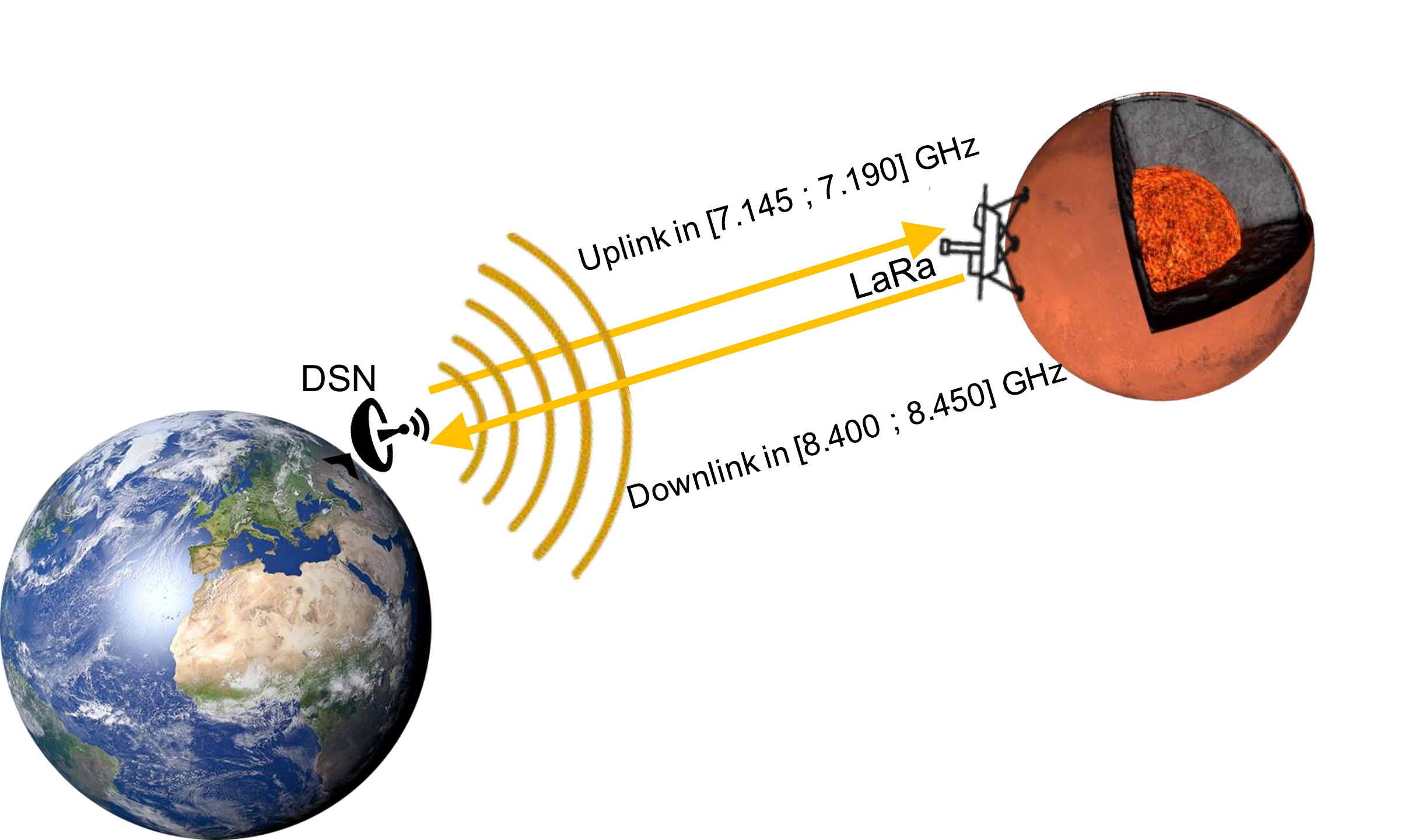}
\caption{Representation of LaRa radiolinks between Earth and Mars.
}
\label{figRadioLink}
\end{figure}

As explained in the introduction, the LaRa experiment uses a coherent X-band 
radio link to obtain two-way Doppler measurements between the Earth and the 
ExoMars lander on Mars. More specifically, measuring the relative position of 
the lander on the surface of Mars with respect to the terrestrial ground 
stations allows reconstructing Mars’ time-varying orientation and rotation in 
space, see Fig.~\ref{figRadioLink}. 
 
\subsection{History}
The LaRa experiment is a legacy of the NEtlander Ionosphere and Geodesy 
Experiment (NEIGE) \citep{Barriot:2001yb,Dehant:2004uf}. The design has changed 
from a three-way link lander-orbiter/orbiter-Earth to a direct-to-Earth (DTE) 
two-way link, and has been proposed to many missions 
\citep{Harri:1999uq,Lognonne:2000aa,Dehant:2009cq,Dehant:2011fk,Dehant:2012uq}. 
At that time, we were planning to use two frequencies, UHF (Ultra High 
Frequency) and X-band. NEIGE was involving tracking from an orbiter and precise 
orbit determination \citep{Yseb03}, while LaRa does not. NetLander was a 
proposal for a network of scientific experiments addressing the Martian 
geophysics. After the rejection of the mission, the experiment was proposed as 
part of the Geophysics and Environmental Package (GEP) for being integrated in 
the platform landing on Mars during the first part of the ExoMars mission to 
Mars in 2016. The complexity and mass budget of the platform then induced a 
rejection of the GEP.\\ 
Several times accepted for a pre-Phase A study, the instrument had gained enough 
maturity to be proposed and to be accepted on the ExoMars 2020 mission to Mars.
The Announcement of Opportunity was issued on 31 March 2015 by the European 
Space Agency (ESA) and the Space Research Institute of the Russian Academy of 
Sciences (IKI RAS) for European payload. The ExoMars 2020 rover and 
\red{Kazachok} Surface Platform (SP) will land on the surface of Mars in March 
2021. The rover will be equipped with a suite of geology and life trace seeking experiments.
After the rover will have egressed, the ExoMars SP, which will contain a further 
suite of instruments, will begin its science mission to study the environmental 
and geophysical nature of the landing site, and among other goals, the 
atmosphere/surface volatile exchange, as well as the geophysical investigations 
of Mars' internal structure providing the general conditions for understanding 
the habitability of Mars at present and in the past. The SP and the rover take 
part of the 2000~kg Descent Module entry mass. \\
\\
In 2015, ESA and IKI issued an Announcement of Opportunity for European payload 
elements on the surface platform of the ExoMars 2020 mission. The selection 
consisted in a technical and programmatic review of the instruments' proposal 
and in a thorough peer review process, performed by a Payload Review Committee, 
consisting of independent scientists, with competences covering the main 
scientific areas of the mission. The decision by the Science Programme Committee 
(SPC) at their meeting on 4-5 November 2015 and by the Programme Board for Human 
Spaceflight, Microgravity and Exploration (PB-HME) at their meeting on 17-18 
November 2015, of ESA Member States selected the European payload elements on 
the Russian surface platform. LaRa, benefiting from a solid heritage (Technology 
Readiness Level 5) built on the design and breadboarding of Orban Microwave 
Product (OMP), was selected as part of this payload because its scientific 
objectives fell within the surface platform Science Priorities of the mission.

\subsection{The teams and institutions}
The experiment is in the hands of the Principal Investigator (PI), the 
scientific team, the instrument team and the Belgian authorities.
The scientific team around the PI, V\'eronique Dehant, is mainly from Belgium, 
France, Russia, the Netherlands and the United States.

The instrument team is a consortium involving the Royal Observatory of Belgium 
(ROB) PI team, industries, and the space agencies. The main role of the ROB team 
in the instrument development is to ensure that the scientific requirements are 
respected and the science objectives will be met. The OHB company Antwerp Space 
is the prime contractor of LaRa, responsible for the overall project and 
especially for the design and manufacturing of the electronic box of LaRa. The 
Universit\'e Catholique de Louvain (UCLouvain) designed the three antennas of 
LaRa. A series of subcontractors are also involved in the LaRa design, 
manufacturing and testing. The work is financially supported by the Belgian 
PRODEX (PROgramme for the Development of scientific EXperiments) program managed 
by ESA in collaboration with the Belgian Federal Science Policy Office 
(BELSPO). 
The ESA PRODEX office is closely following the LaRa project, from both 
managerial and technical aspects. Besides this consortium in charge of 
developing the instrument, BELSPO and the ESA PRODEX support the ROB team for 
LaRa operations and ground support. Together with ROB, ESOC (European Space 
Operations Centre) is involved in the preparation of the ground segment for the 
LaRa operations and will conduct the operations of LaRa for the entire mission 
duration. As prime of the ExoMars 2020 \red{Kazachok} Surface Platform and 
responsible for the science operations of it, the Space Research Institute of 
the Russian Academy of Sciences (IKI), the Lavochkin Scientific and Production 
Association and the Roscosmos State Corporation of Russia are important actors 
of the LaRa project. NASA will also be part of the LaRa adventure by providing 
tracking time on NASA’s Deep Space Network (DSN) \red{at the level of two 
one-hour pass per week} as part of an agreement between NASA and ESA.   
Supplementary measurements using Planetary Radio Interferometry and Doppler 
Experiment (PRIDE) technique in support to LaRa will be conducted by JIVE, TU 
Delft in cooperation with the European VLBI Network (EVN) and other radio 
astronomy observatories around the world. BELSPO funds 100\% of the LaRa 
instrument, except for the DSN and PRIDE support.

\subsection{The transponder}
\begin{figure}[!ht]
\centering
\includegraphics[width=0.6\textwidth]{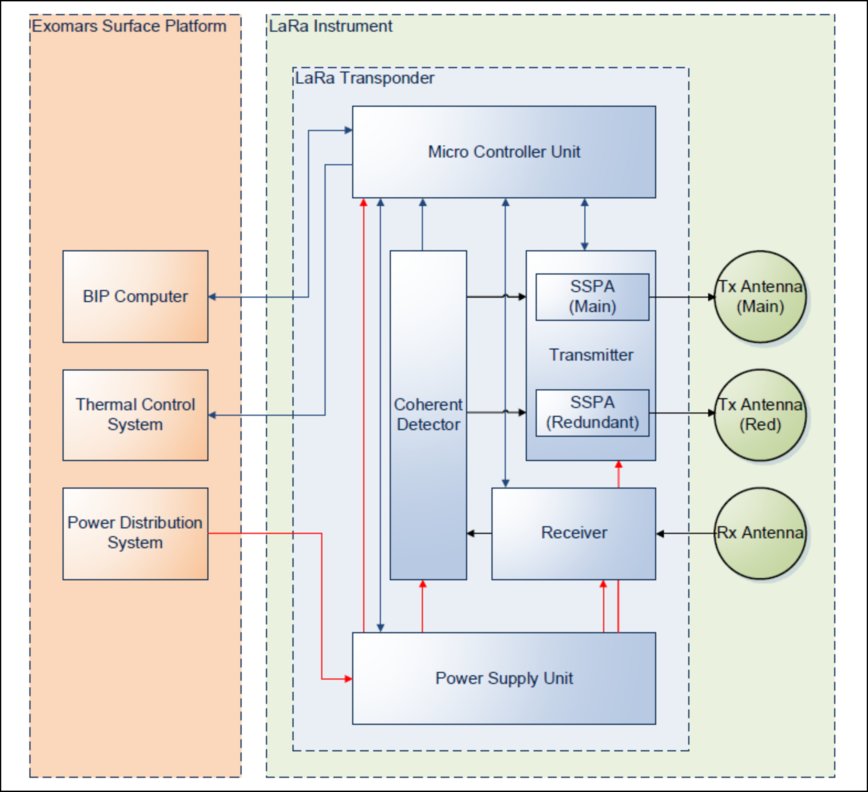}
\caption{LaRa design of the transponder.}
\label{figLaRainterior} 
\end{figure}

The main element of the LaRa onboard instrumentation is the 
$25\times8\times8$~cm electronic transponder box. This box weights 1.5~kg 
\red{while the total weight of LaRa is 2.15~kg}. The transponder has two 
functional modes: a functioning mode referred to as a power-on mode and a 
sleeping mode referred to as a power-off mode. When turned on, the LaRa 
instrument uses at most 42~W (nominal power provided by the platform) to produce 
a radio frequency output power of about 5~W. 
LaRa will be turned on a few minutes before the expected arrival time of the 
uplink radio signal sent by the ground station. 
The receiver of the X-band RF transponder consists of an narrow-band input 
band-pass filter, a three-stage low-noise amplifier followed by frequency 
down-converters that split the required Rx gain of more than 140dB over two 
down-converting stages. The second intermediate frequency output is then fed to 
a coherent detector that compares it continuously with an internally generated 
crystal reference clock such that the output signal error is minimized leading 
to uplink signal acquisition and precise Doppler tracking. 
The signal enters then the heart of the LaRa transponder, namely the coherent 
detector. 
The locked signal at internal frequency is then injected in the transponder 
transmitter chain, which will up-convert it to the X-band output frequency. 
Finally, a Solid-State Power Amplifier (SSPA) amplifies the signal that will be 
radiated by one of the transmitting antennas (see details below). This process 
ensures that the output signal is coherent with the incoming signal although not 
at the exactly same frequency because of the 880/749 turn around ratio applied 
to it (as imposed by ITU, International Telecommunication Union, regulations for 
deep space missions \citep{ITU-ch24}). Without a signal present at the receiver 
input port, the crystal oscillator is swept by a ramp generator. The transponder 
operates in non-coherent mode transmitting a continuous wave signal. In nominal 
mode, LaRa can transmit coherently (locked) and non-coherently (unlocked). The 
transponder contains a micro-controller unit handling the telemetry (LaRa health 
data) and the transponder electronics. For the sake of risk minimization, the 
SSPA and transmitting antenna (Tx) are redundant. The transponder is connected 
to the power distribution system, the thermal control system, and the SP onboard 
computer (so-called BIP computer). The complete design is shown in 
Fig.~\ref{figLaRainterior}.

\begin{figure}[!ht]
\centering
\includegraphics[width=0.8\textwidth]{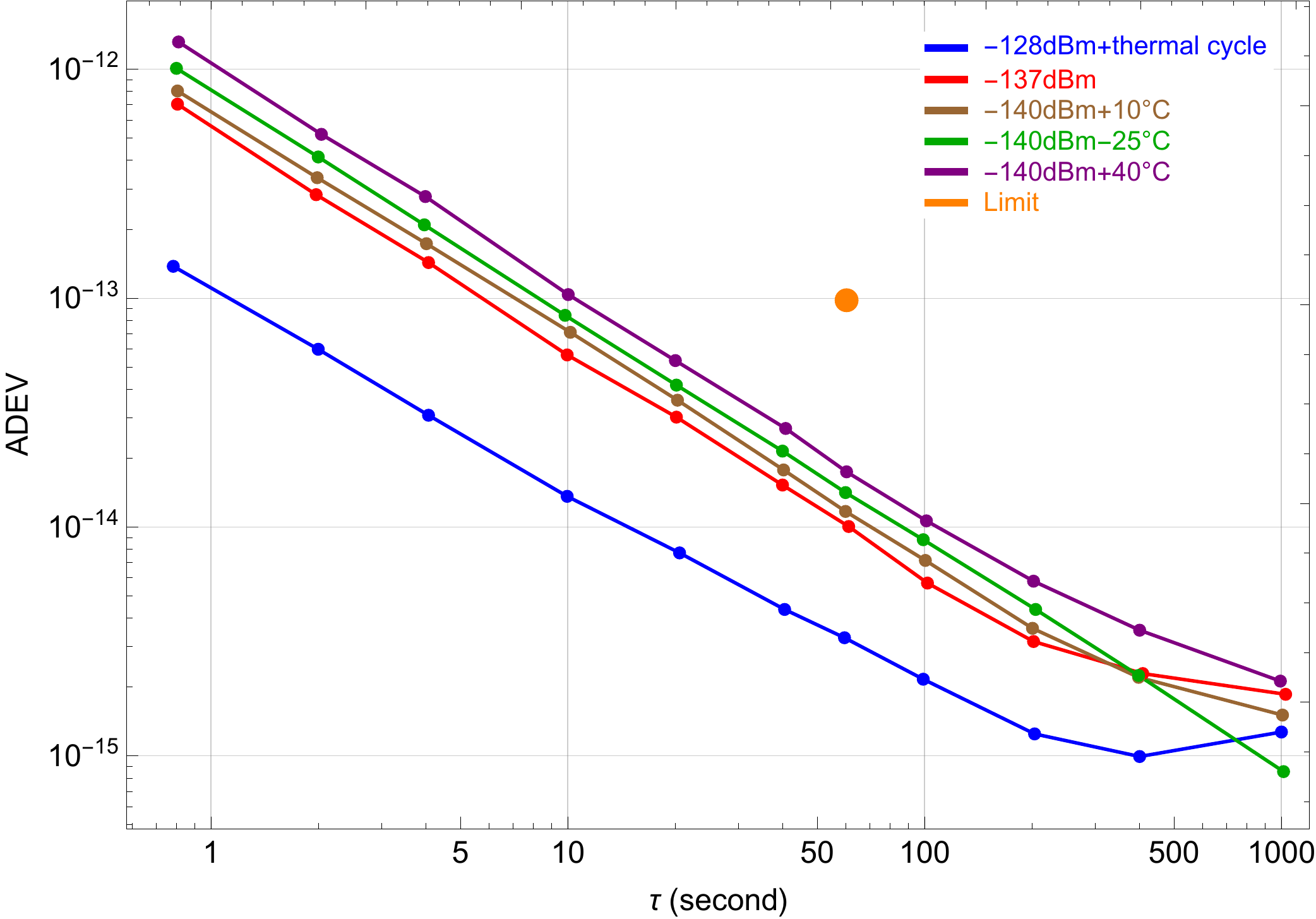}
\caption{Allan deviation (ADEV) of the LaRa instrument computed when running at 
different temperatures and for different input power levels. Thermal cycle 
(blue) consists in a temperature variation at a rate of 10$^\circ$C/h. The 
orange dot (called ''limit'') represents the requirement ($10^{-13}$ at 60 
seconds integration time).}
\label{figadev}
\end{figure}

The design of the LaRa transponder has been finalized at the Critical Design 
Review (CDR) held in October 2018. The performance of the instrument has been 
demonstrated to be compliant with the scientific requirements. In particular, 
the frequency stability has been measured at different input power level (down 
to $-148$~dBm) and at different temperatures. The temperature range considered 
for the tests is driven by the fact that the LaRa electronics/transponder will 
be maintained at a temperature between $-20^\circ$C and $+40^\circ$C since it 
will be controlled by the RHU (Radioisotope Heater Unit) of the Surface 
Platform. The frequency stability is commonly characterized by the Allan 
deviation of the signal over an observation time $\tau$. As shown in 
Fig.~\ref{figadev}, the Allan deviation of all measurements are below the 
requirement of $10^{-13}$ at 60s integration time.

The functionality of LaRa has been validated using Electrical Ground Support 
Equipment (EGSE) facility of ESOC (see Fig.~\ref{figEMtest}).

The LaRa model list allowing to meet scheduling and funding conditions is the 
following:
\begin{itemize}
\item A Structural Model (SM) for mechanical verification,
\item A Thermal Model (TM) for thermal verification,
\item An Electrical Interface Simulator (EIS) for electrical interfaces for SP 
interface verification,
\item A Qualification shock model,
\item Two Engineering Models (EM) for qualification and test such as radio 
frequency performance, SP interfaces, as well as for on-ground test model after 
launch,
\item A Proto Flight Model (PFM) for flight that will undergo acceptance testing 
and some verification testing and minimum qualification testing,
\item A Flight Spare kit (FS kit) consisting of PFM duplicate electronic 
components, and
\item A Ground Support Equipment (GSE).
\end{itemize}
Some of these models and their testing are shown in 
Fig.~\ref{figTransponderTest}, \ref{figAntennaTest}, and 
\ref{figTransponderAntennaTVAC}.

\subsection{The monopole antennas}
\begin{figure}[!ht]
\centering
\includegraphics[width=0.3 \textwidth]{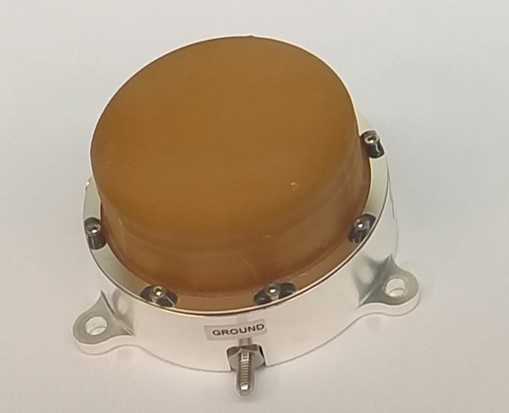}
\includegraphics[width=0.28\textwidth]{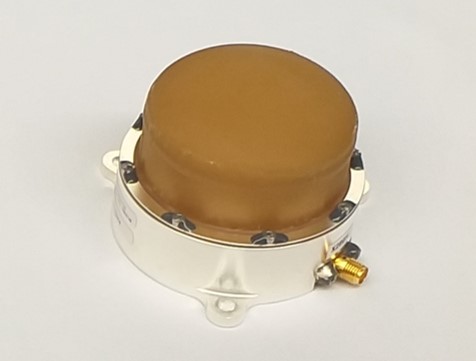}
\includegraphics[width=0.35\textwidth]{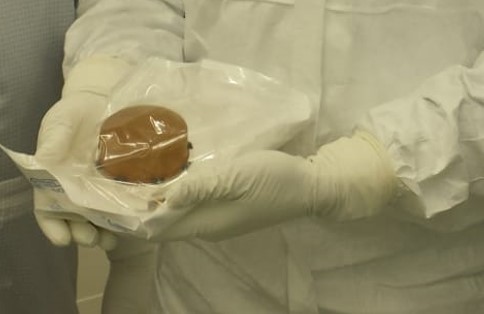}
\caption{Pictures of the receiving (Rx) (left) and transmitting (Tx) antennas 
(center), and Tx when ready for pattern measurement, after planetary protection 
process (right).}
\label{figantenna}
\end{figure}

\begin{figure}[!ht]
\centering
\includegraphics[width=0.39\textwidth]{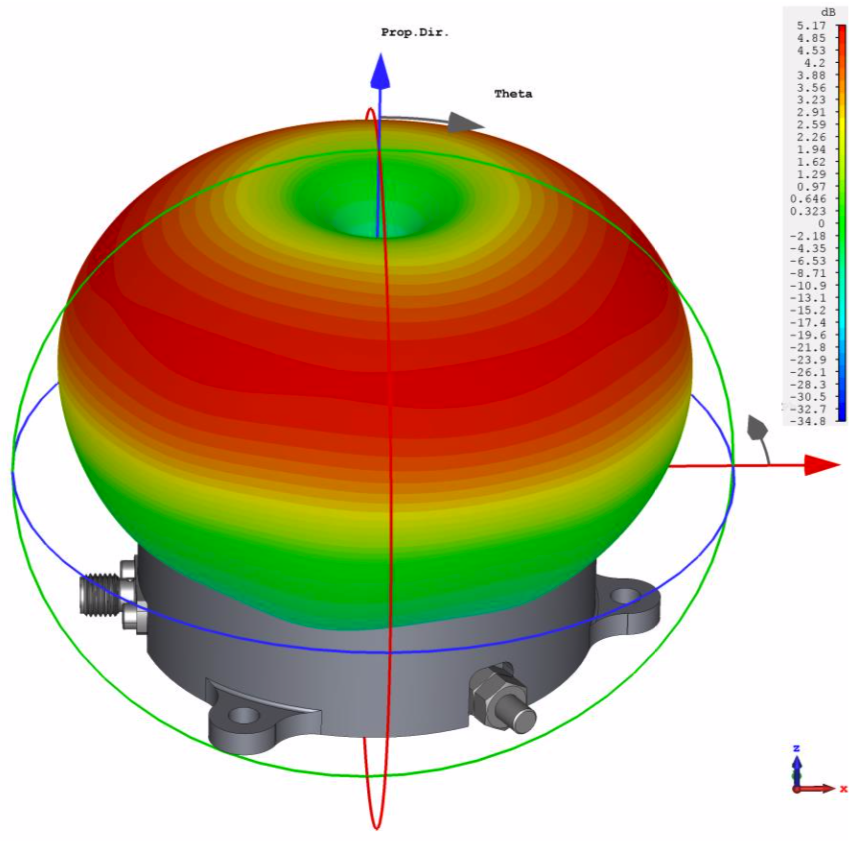}
\includegraphics[width=0.42\textwidth]{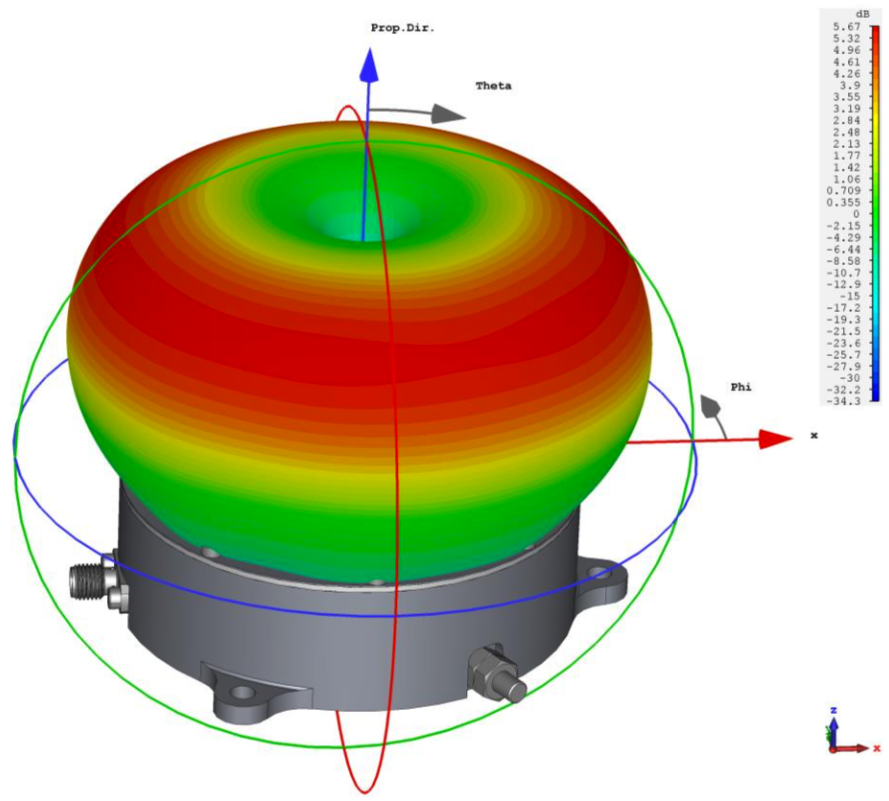}
\caption{Computed radiation Pattern (3D) of the Rx (left) and Tx (right) 
antennas, above aluminium housing.}
\label{fig3Dpattern}
\end{figure}

\begin{figure}[!ht]
\hspace*{-1cm}
\includegraphics[width=0.67\textwidth]{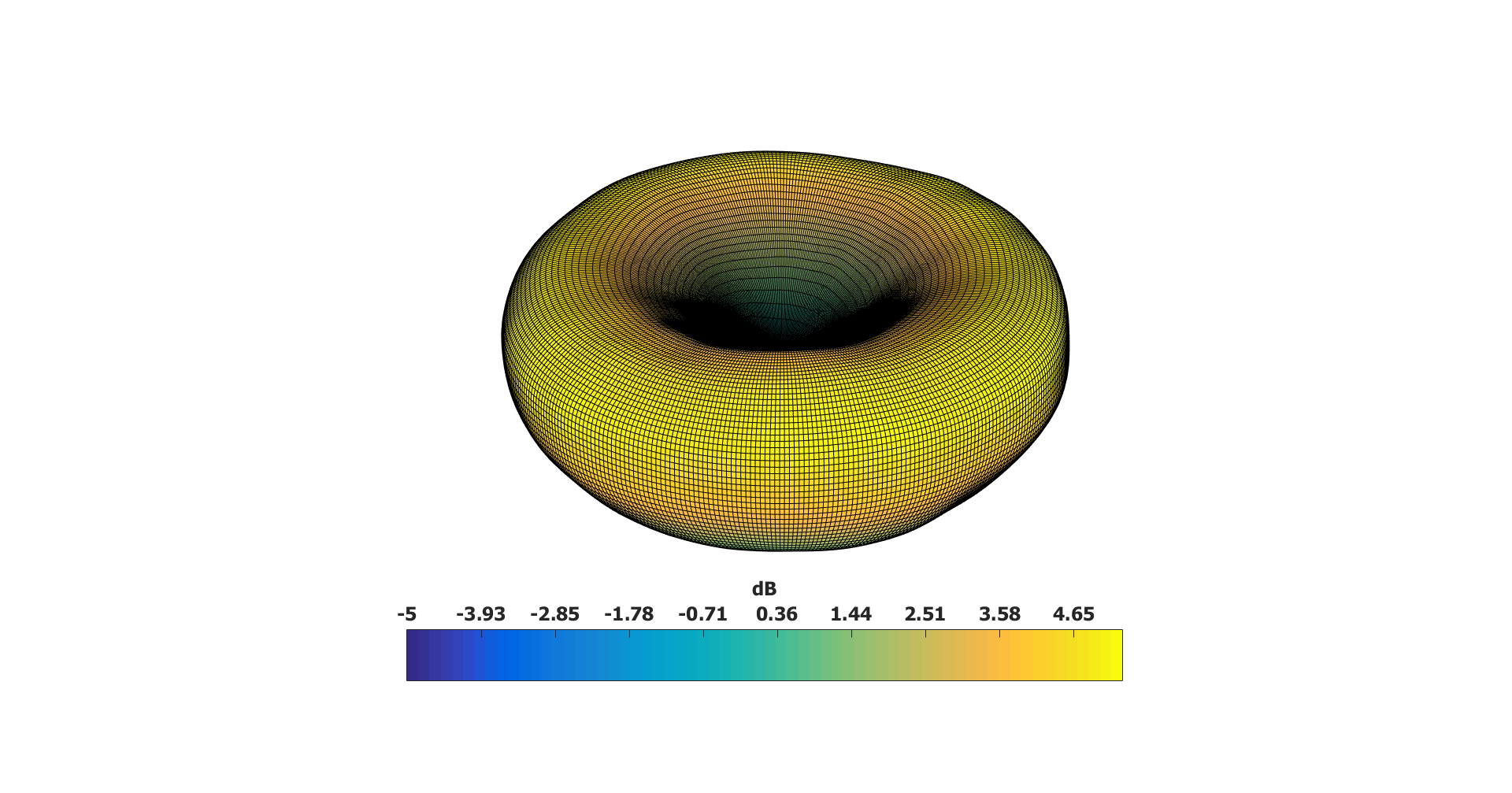}
\includegraphics[width=0.33\textwidth]{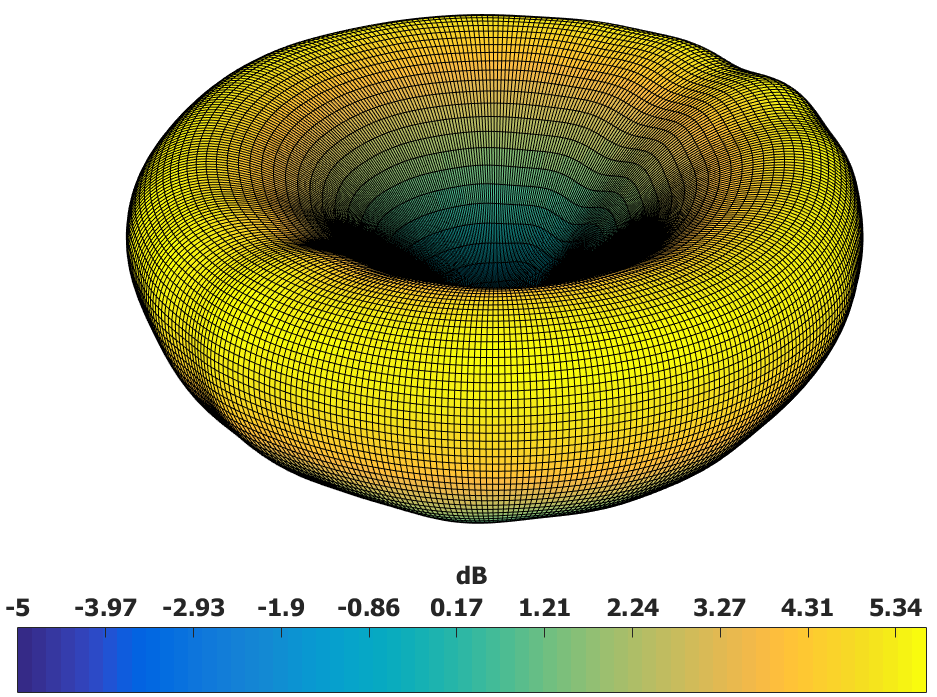}
\caption{Observed radiation Pattern (3D) of the Rx (left) and Tx 
(right) antennas. Small ripples in the Tx pattern are probably due to scattering 
by a feeding cable}
\label{fig3DOBSpattern}
\end{figure}

\begin{figure}[!ht]
\centering
\includegraphics[width=0.4\textwidth]{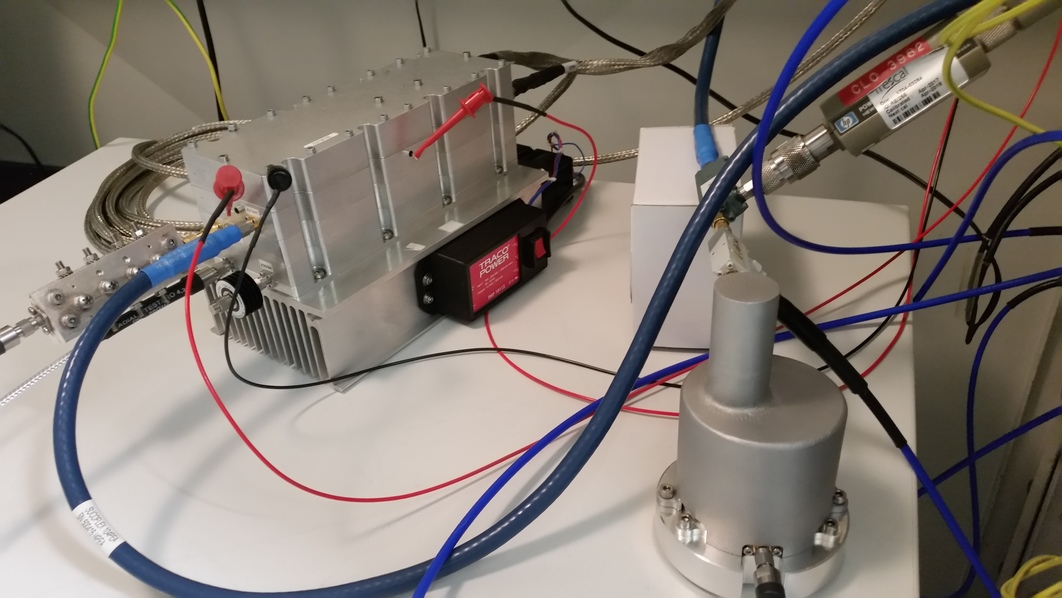}
\caption{Photo of the end-to-end testing at ESOC (European Space Operations 
Centre, in Darmstadt, Germany) of the Engineering Models of the Electronic box 
(E-box, transponder part) and test caps containing the antennas. }
\label{figEMtest}
\end{figure}

\begin{figure}[!ht]
\centering
\includegraphics[width=0.3\textwidth]{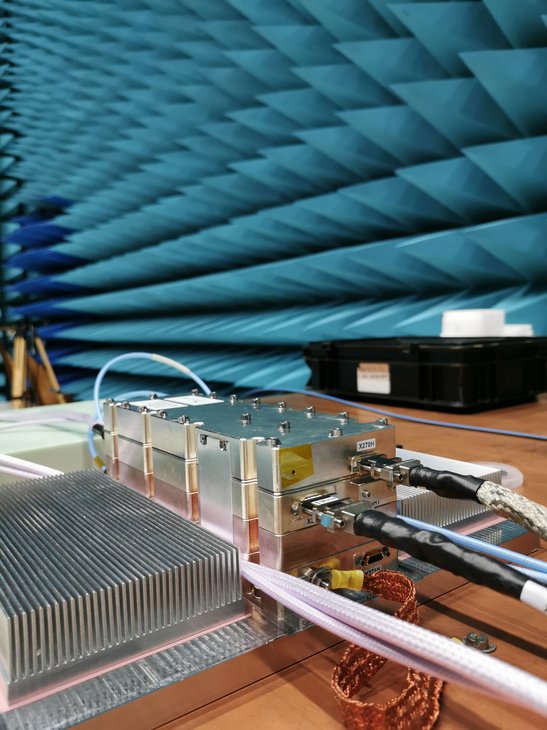}
\includegraphics[width=0.3\textwidth]{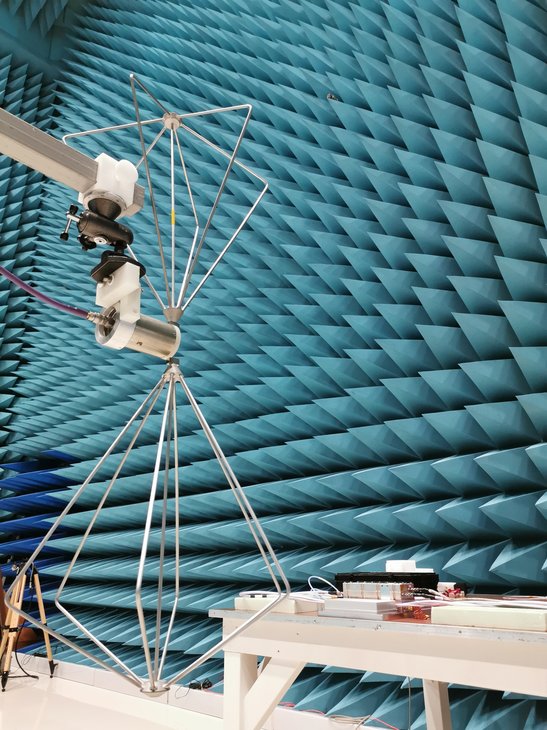}
\includegraphics[width=0.5\textwidth]{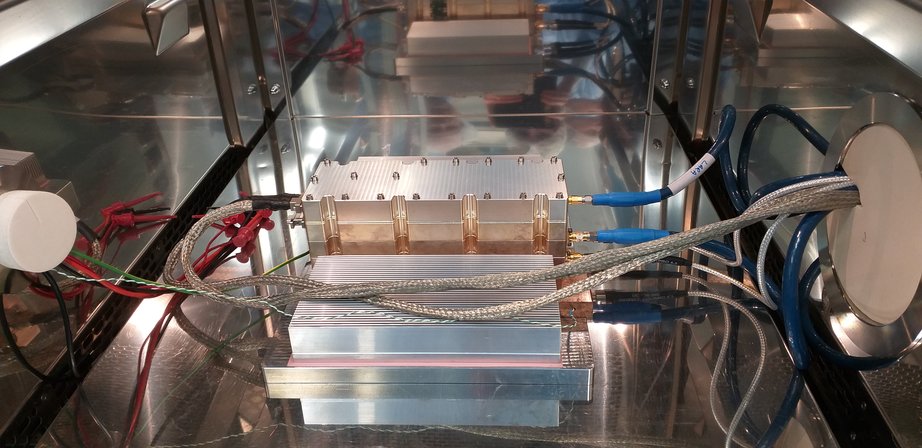}
\includegraphics[width=0.4\textwidth]{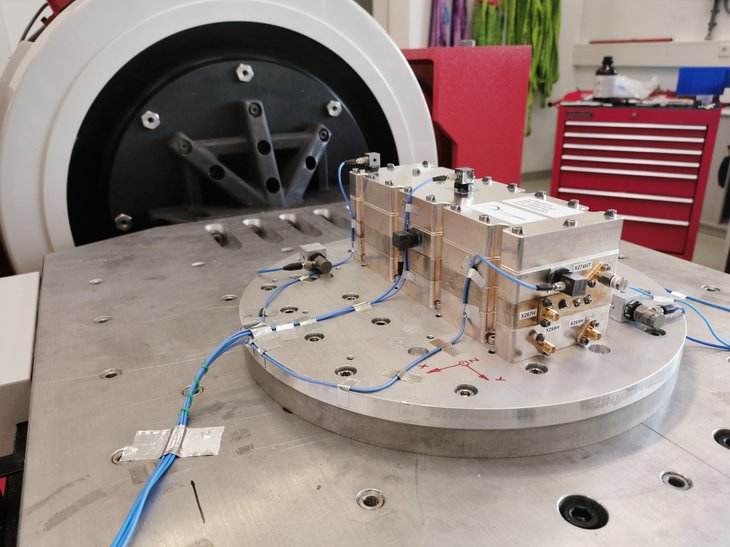}
\includegraphics[width=0.4\textwidth]{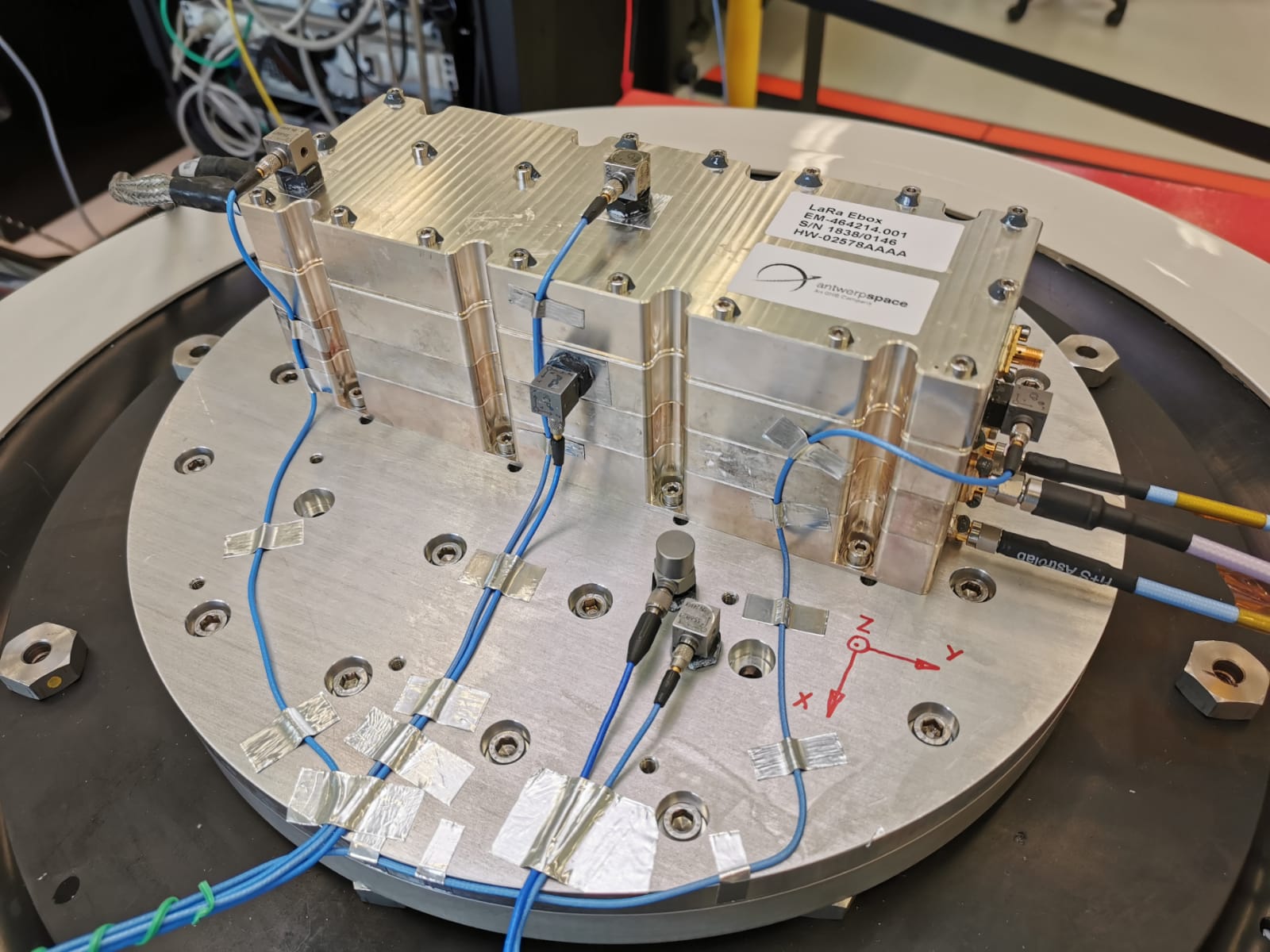}
\caption{Top-left: flight model of the transponder during EMC testing. 
Top-right: EMC facility. Medium-left: flight model of the transponder in a 
thermally-controlled chamber when testing at AntwerpSpace; the e-box is in the 
back of the chamber. Medium-right: flight model of the transponder during 
vibration tests. Bottom: flight model of the transponder just before entering 
TVAC chamber.}
\label{figTransponderTest}
\end{figure}

\begin{figure}[!ht]
\hspace{2.3cm}
\includegraphics[width=0.5\textwidth, angle=-90]{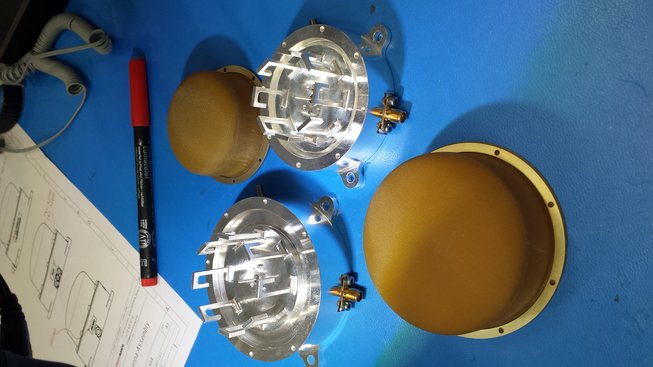}

\vspace{-5cm}
\hspace{6.7cm} 
\includegraphics[width=0.4\textwidth]{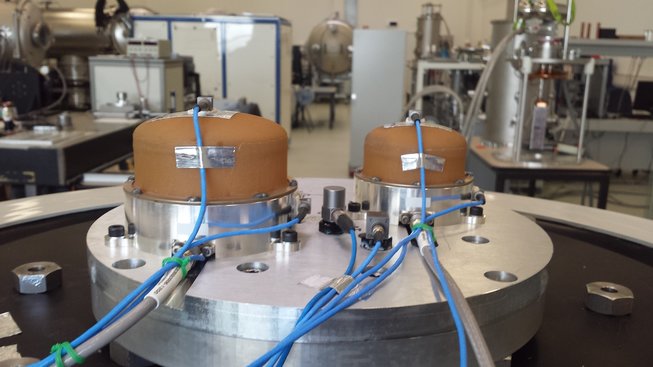}

\centering
\vspace{1.8cm}
\includegraphics[width=0.4\textwidth]{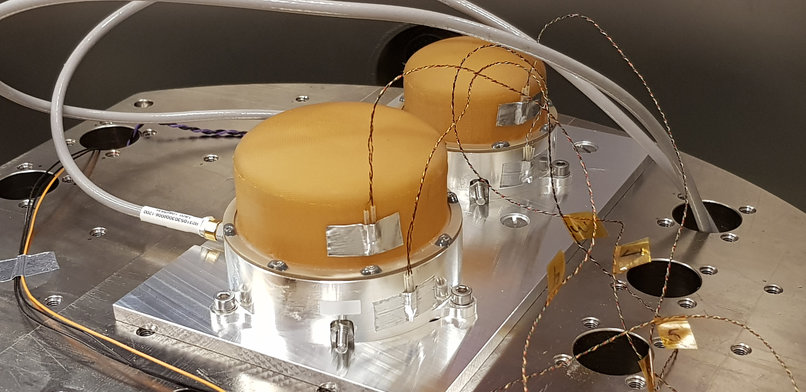}
\includegraphics[width=0.4\textwidth]{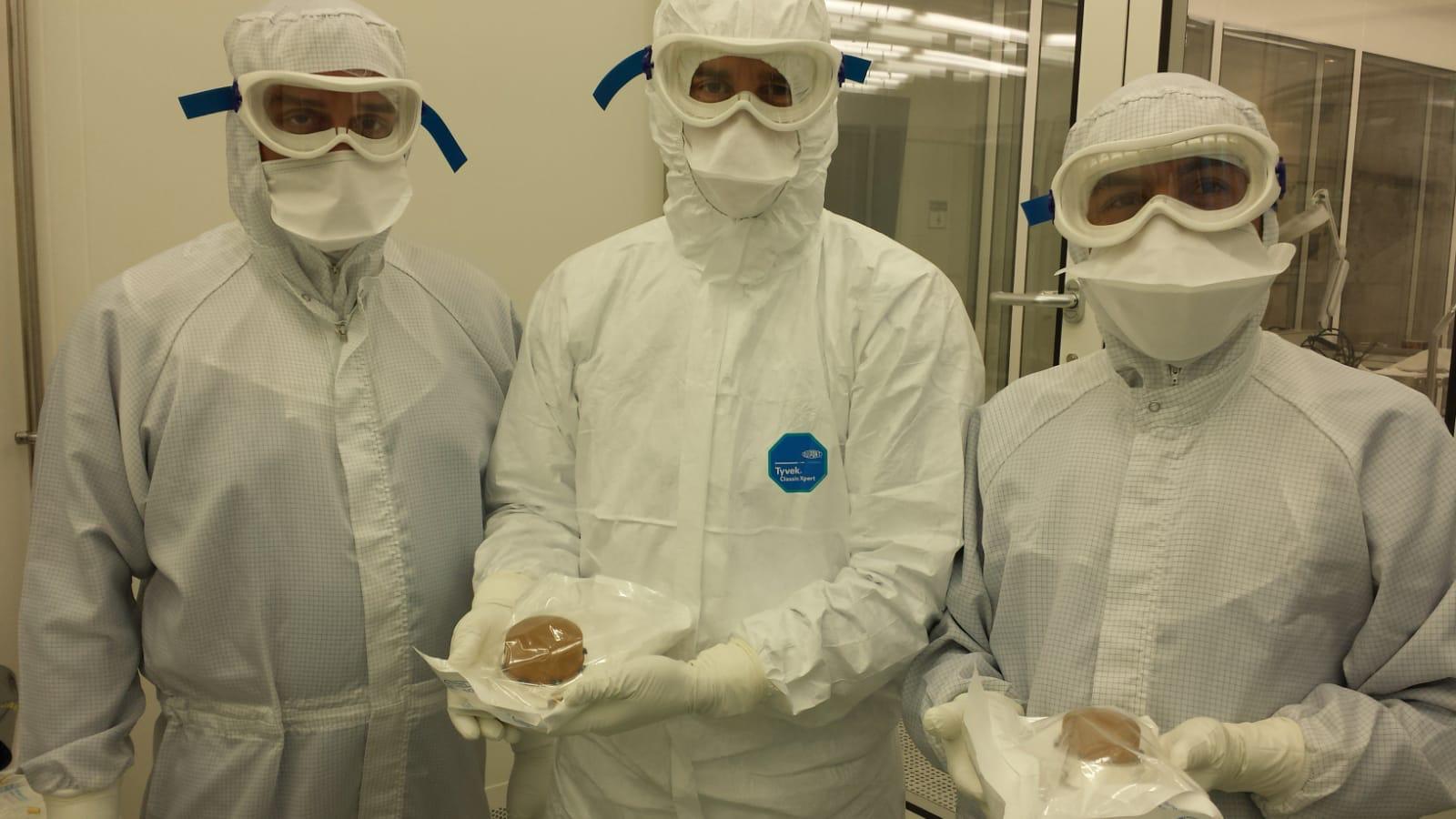}
\caption{Top-left: flight model of the antennas open to show the parasitic form 
of the antennas. Top-right: flight models of antennas during vibration tests. 
Bottom-right: flight models of antennas before entering TVAC chamber. 
Bottom-right: flight models of antennas in the hands of the responsible 
engineers. }
\label{figAntennaTest}
\end{figure}

\begin{figure}[!ht]
\centering
\includegraphics[width=0.4\textwidth]{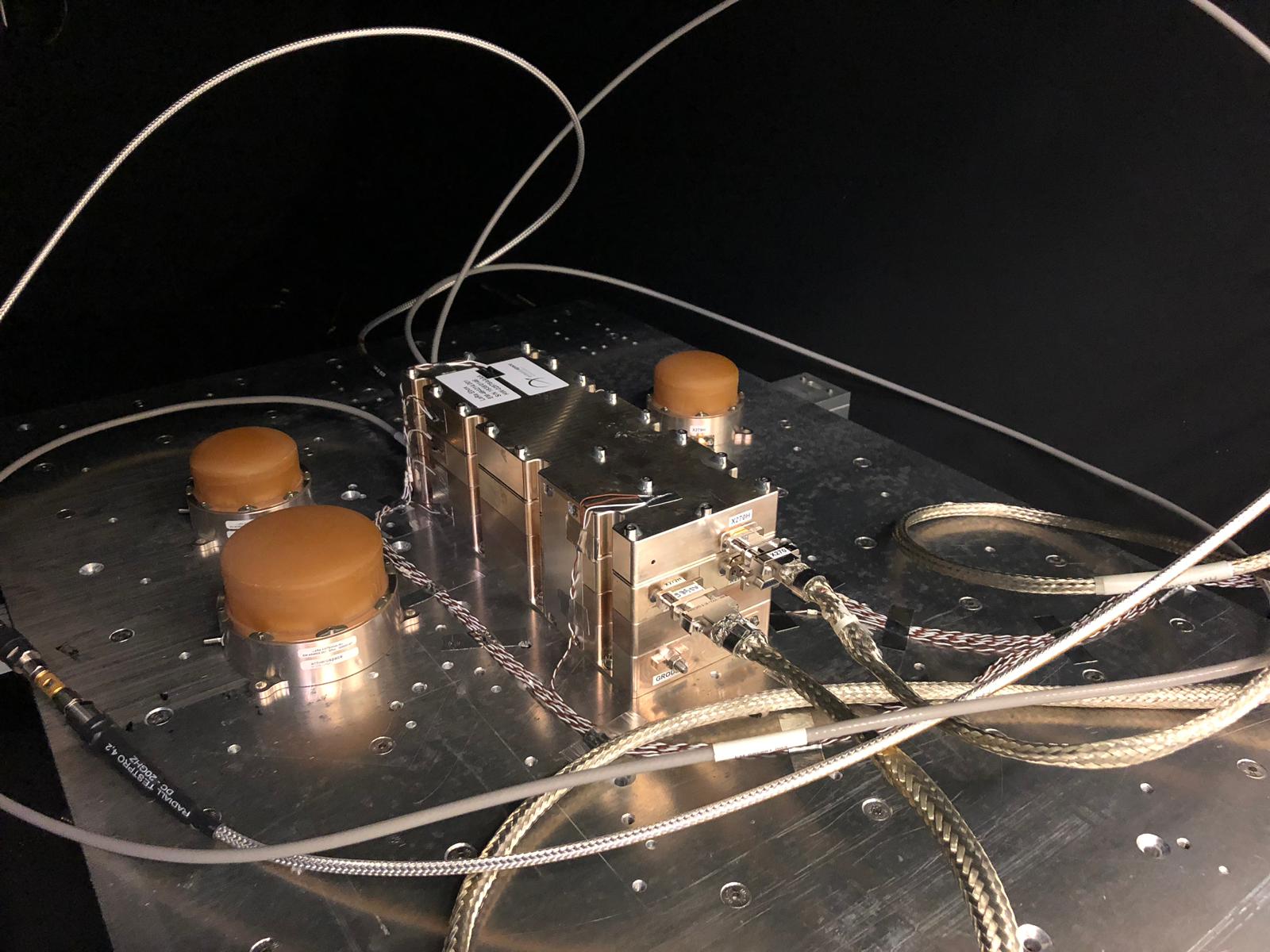}
\includegraphics[width=0.4\textwidth]{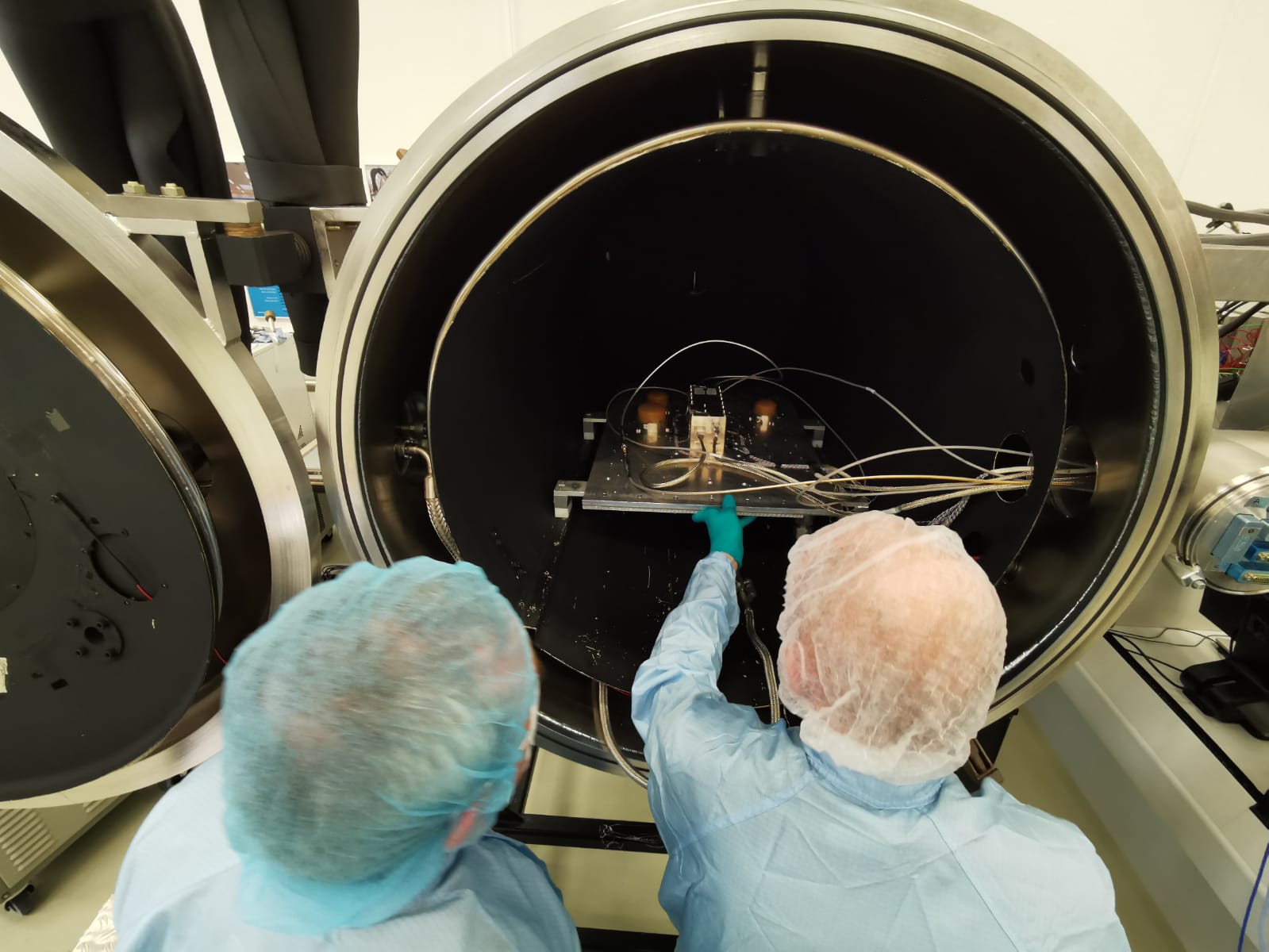}
\caption{Photos of the flight models of the antennas and transponder entering 
the TVAC testing. }
\label{figTransponderAntennaTVAC}
\end{figure}

LaRa will use one antenna for receiving the signal from the Earth and two 
antennas (to ensure redundancy \red{of the SSPA just prior to the transmitting 
antennas in the transponder output chain}) for retransmitting the signal back to 
Earth (see detailed description of LaRa's antennas in \citet{Karki:2019aa}). The 
transmitting (Tx) and receiving (Rx) antennas (see Fig.~\ref{figantenna}) 
operate in X-band (channel 24) around 8428.580248~MHz and 7173.871143~MHz, 
respectively. They radiate conical patterns with a maximum gain of about 5~dBi 
(antenna gain in dB w.r.t.~an isotropic radiator) and a main lobe in the 
[30$^\circ$, 55$^\circ$] range of elevation, see Fig.~\ref{fig3Dpattern}, with 
right-hand circular polarization. This allows a good link budget with the Earth, 
over its path in Martian sky. The observed effective patterns are shown in 
Fig.~\ref{fig3DOBSpattern} (see also \citet{Karki:2019aa}).
Gamma-shaped parasitic elements surround the centrally fed monopole. They 
protrude from an aluminum housing, which also includes a choke ring, intended to 
cut edge currents and reduce the back-lobes (see Fig.~\ref{figAntennaTest},  
top-left). The Tx and Rx antenna diameters are 66 and 80~mm, and their masses 
are 132 and 162~g, respectively. A radome also covers the whole structure. The 
radiator itself is made of a monopole fed through a square-coaxial transmission 
line. The latter is fed from the side of the aluminum housing. The shape of the 
parasitic elements controls the circular polarization purity. \\

Unlike the LaRa electronic box, the antennas will not be mounted on a thermally 
controlled panel and will therefore experience a larger range of environmental 
temperatures, between $-120^\circ$C and $+60^\circ$C. As a result, a dedicated 
qualification campaign for materials and process used for these antennas had to 
be carried out. The fully metallic design of the radiator itself enables 
withstanding the huge temperature range.

The RF harness, i.e.~coaxial cables linking the antennas to the transponder box, 
are about 2~m long, inducing a 1.7~dB power loss in the link-budget (see below).

\subsection{The ground segment}
The LaRa experiment involves a space segment, which is the LaRa instrument 
itself (i.e.,~transponder and receiving/transmitting antennas), as well as a 
ground segment consisting in three potential networks of large antennas 
distributed around the globe. The chosen nominal one is the NASA Deep Space 
Network (DSN) that consists of three deep-space communications facilities placed 
approximately 120 degrees apart in longitude: (1) at Goldstone, in California's 
Mojave Desert, USA; (2) at Robledo near Madrid, Spain; and (3) at Tidbinbilla 
near Canberra, Australia. Each of these facilities includes several 34~m 
antennas and one 70~m antenna. The 70~m DSN antennas are baseline for LaRa 
ground segment as only these are currently able to track the weak signal 
transponded by LaRa during a whole Martian year (see next subsections). 

As backup solution, when the Earth-Mars distance is basically smaller than 
1.5--2 astronomical units (AU), the LaRa experiment foresees to also use the 
35~m antennas of the ESA's ESTRACK network. ESTRACK stations are located in 
Cebreros near Madrid, Spain; in New Norcia, Australia; and near Malarg\'{u}e, 
Argentina.

The Russian Ground Station (RGS) network might also be used to track LaRa 
provided some major upgrades of the RGS antennas are performed. The network 
consists of two 64~m antennas in Bear Lakes near Moscow and Kalyazin in Tver 
area, and a 70~m antenna in Ussuriisk, north of Vladivostok. Unlike the other 
two networks, the Russian antennas are all located in the northern hemisphere 
within a limited range of longitudes of about 90$^\circ$, which makes them much 
less convenient for the experiment during an entire Martian year. Indeed Mars 
barely rises above 10º elevation in the RGS sky during several consecutive weeks 
(about 30 weeks below 30º elevation) every Martian year. As shown in the next 
section, this has a strong impact on the link budget because of the large 
atmospheric losses at low elevation angle and the absence of global coverage in 
the antenna positions (only in Russia). For these reasons, these ground stations 
are disregarded at this point of the project. 

All ground stations mentioned above are equipped with ultra-stable hydrogen 
maser local oscillators (LO) ensuring the highest reference frequency stability.

In addition to the space agencies' deep space tracking networks, a global 
network of radio telescopes equipped with Very Long Baseline Interferometry 
(VLBI) instrumentation has the potential of providing useful measurements for 
the LaRa experiment. About forty VLBI radio telescope antennas ranging in 
diameters from 220~m (the illuminated diameter of the Arecibo radio telescope) 
to about 20~m are distributed over all continents. These telescopes can operate 
using the PRIDE technique for obtaining Doppler measurements for LaRa in both 
the nominal two-way coherent regime and in the non-coherent free-running regime. 
The former case represents a so called 'three-way' configuration (up- and 
downlink from/to a ground station plus a one-way downlink from SP to a radio 
telescope) or a one-way -- just a downlink to a radio telescope in a 
free-running mode. 
\subsection{The link budget}
\begin{figure}[!ht]
\hspace{-0.4cm}
\includegraphics[width=0.53\textwidth]{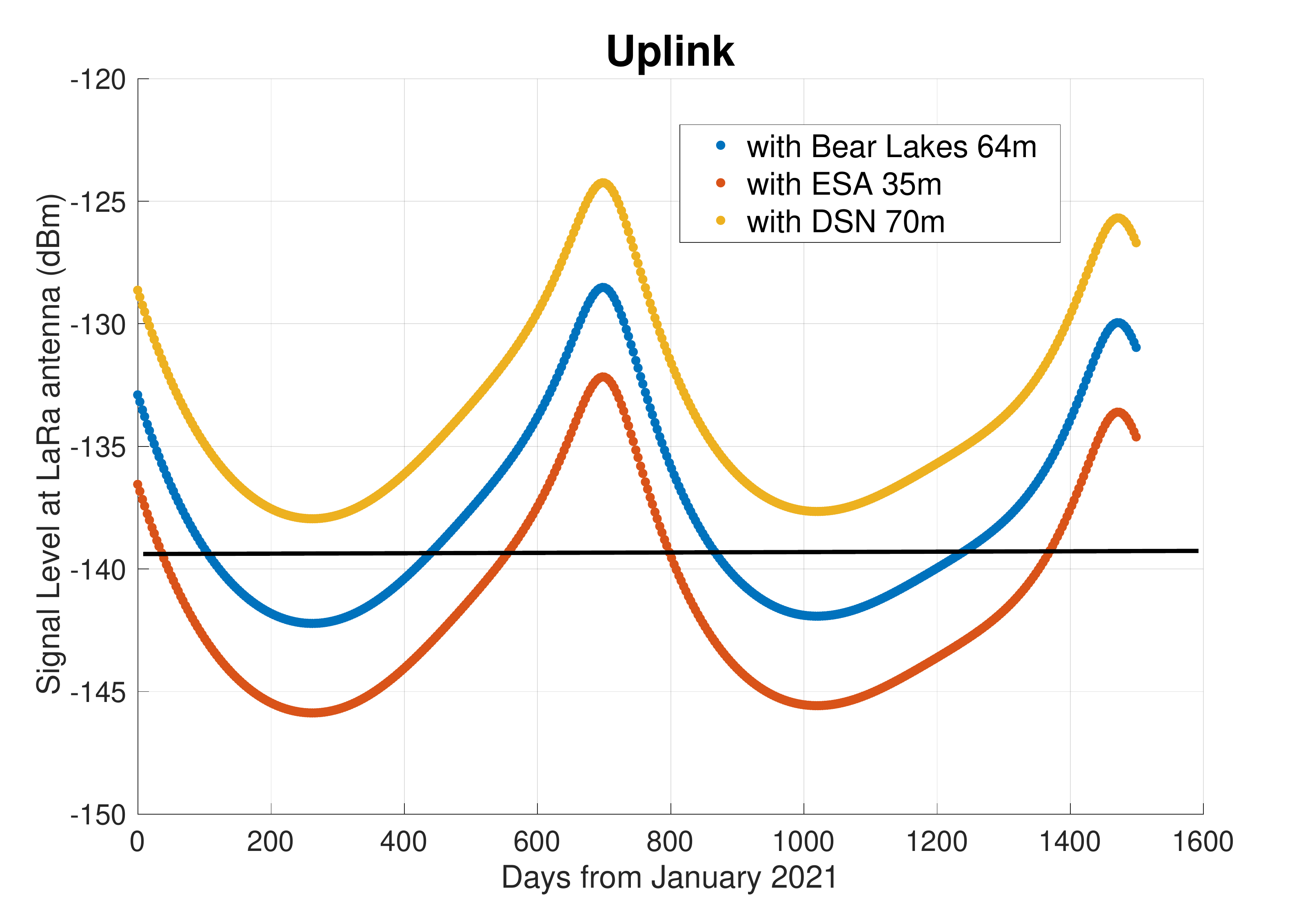}
\includegraphics[width=0.53\textwidth]{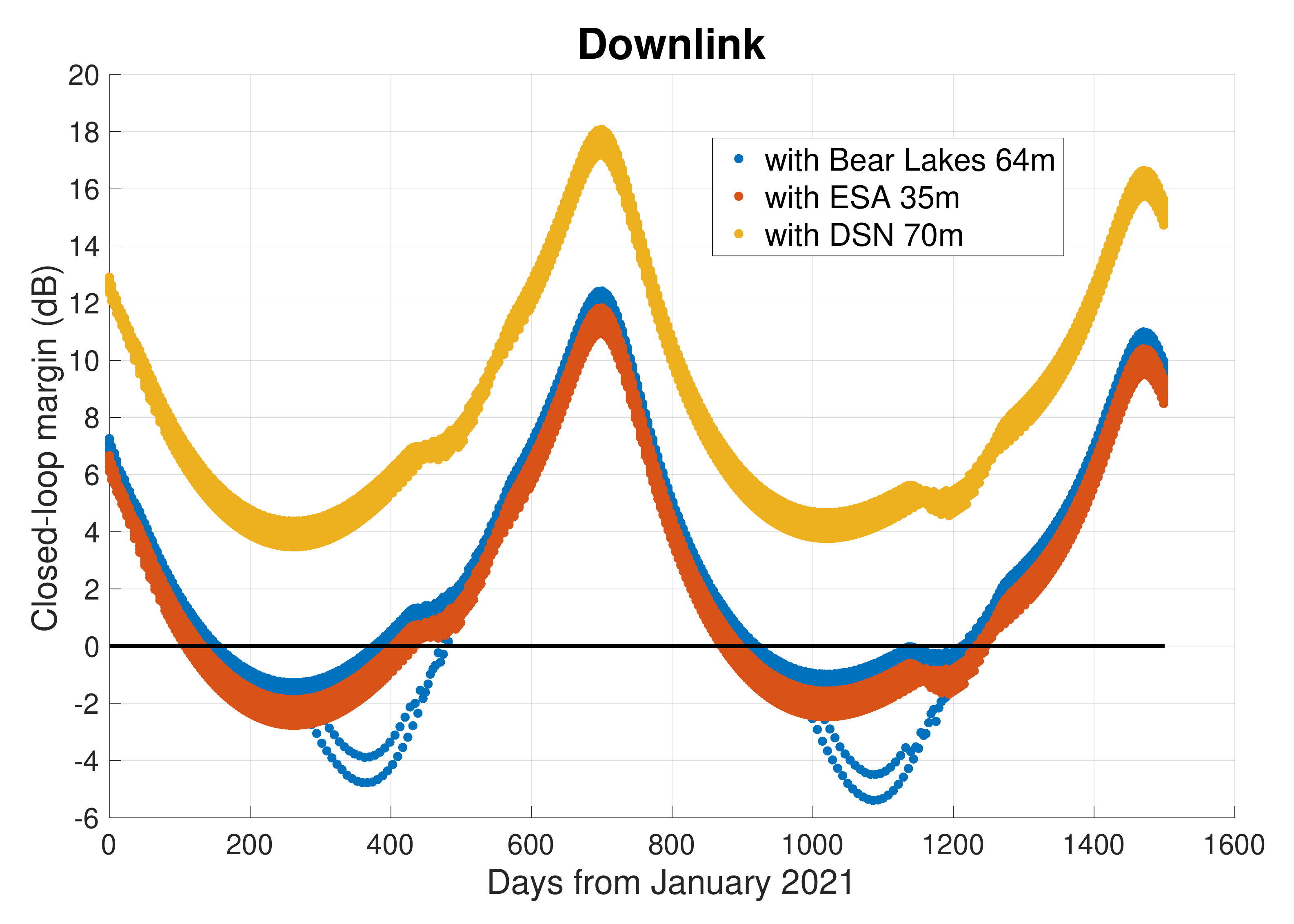}
\caption{LaRa uplink signal level at Rx antenna (left) (which includes cable 
loss and antenna gain) and SNR margin for closed-loop tracking of LaRa's 
downlink signal (right). The black lines indicate the lower limit above which 
the signal can be locked at the LaRa antenna (left) or at the ground station 
(right).}
\label{fig:links}
\end{figure}

For the design of a radioscience experiment like LaRa, the link budget is an 
important aspect to be analysed in order to ensure the acquisition of Doppler 
measurements at the Earth ground stations. Based on the performance of the LaRa 
instrument itself, and on the performance of the receiving and transmitting 
ground stations, the uplink signal level received at the LaRa receiving antenna 
(Rx) and the signal-to-noise ratio (SNR) expected at the Earth stations can be 
computed, while accounting for the propagation medium in the frequency band of 
interest (X-band). These two quantities are shown in Fig.~\ref{fig:links} for 
four Earth years as a function of time or equivalently Earth-Mars distance, and 
for the different ground stations.

The LaRa transponder has been designed to lock and maintain the lock for more 
than one hour on an uplink signal level down to -140~dBm. 
As shown on the left panel of Fig.~\ref{fig:links}, the LaRa receiver chain 
ensures the tracking of the DSN-70~m uplink signal at any time during the next 
two Martian years. Fig.~\ref{fig:links} right panel, shows the SNR margin at the 
ground station receiver inputs for the 70~m DSN antennas (orange), the 35~m 
ESTRACK antennas (red) and the 64~m RGS antennas (blue). When the curves are 
below zero, the lock can not be maintained anymore in the closed-loop mode and, 
if the uplink is guaranteed, tracking in open-loop mode has to be performed, 
i.e.~instead of using the ground station receiver PLL \red{(Phase-Locked Loop)} 
that directly provides the Doppler, we need to record the whole signal 
\red{returning back from Mars} in a certain frequency band \red{after being 
transponded at Mars with the classical deep-space transponder ratio (880/749)} 
and reconstruct the Doppler afterwards (see also Section~\ref{GSPLL}). As 
clearly shown on this panel, DSN antennas (baseline for LaRa) can always track 
LaRa's signal while the other two can only be used below a certain distance 
(1.5-2~AU for both RGS and ESTRACK stations). The RGS computation has been 
performed assuming Bear Lakes antenna performance and location. As mentioned in 
the previous section, the SNR drops significantly when Mars stays at elevation 
lower than 30º in Bear Lakes' sky due to atmospheric losses (see large 
deflections in blue curve of Fig.~\ref{fig:links} right panel around 350 and 
1100 days). In addition, the link-budget is computed in a worst-case scenario 
where the SP would be tilted 20º Northwards. Then, the Earth would sometimes not 
rise above 25º-30º in LaRa's sky, forcing to track at lower elevation at Mars 
and therefore with lower LaRa antennas gain as suggested by 
Fig.~\ref{fig3Dpattern}. This is responsible for the slight distortions observed 
in the downlink margins of each of the three stations around 500 days and 1200 
days (see Fig.~\ref{fig:links} right panel).  \\
\\
As a conclusion, based on the link-budget analysis, we demonstrate that the LaRa 
experiment is well designed for an actual measurement recording in closed-loop 
mode with DSN (see Section~\ref{GSPLL} for the complete explanation).

It must be mentioned that, for the downlink (not shown here), it is always 
possible to observe in open-loop mode \red{(still in two-way configuration)} 
with any of the above-mentioned networks, if the uplink is working. The 
limitation of the use of the ESTRACK stations and the Russian ground stations, 
is arising from the uplink budget when the LaRa transponder is not able to lock 
on the signal uplink. If the uplink uses a better emission performing station 
like some of the DSN 34~m antennas \red{(not shown in Fig.~\ref{fig:links}}), 
the link can be closed in the open-loop configuration.

\subsection{The measurements}
The LaRa measurements are two-way Doppler shifts acquired directly at the Earth 
ground stations. No data are stored onboard the platform except for a few LaRa 
health measurements such as internal temperatures and voltages of different 
components. These data are transmitted to Earth via the telemetry of the 
platform. 

\subsubsection{Data types }
\paragraph{Ground Station Doppler data}
\label{GSPLL}
The ground tracking stations are equipped with PLL receivers making the tracking 
of Doppler-induced changes in frequency of the X-band downlink received carrier 
possible. These receivers generate a model of the received signal and 
cross-correlate the model with samples of the incoming downlink signal. They use 
a PLL to estimate the difference; that PLL estimate plus the model is the 
measurement. 
PLL receivers allow signal acquisition, lock-up, and detection in real time 
\citep{Morabito:1994fk}.
The closed-loop data are delivered in various format ({\it e.g.} ATDF, ODF, TNF, 
TDM, etc.). Among the quantities on a tracking file are measurement acquisition 
time, Doppler shift, Doppler counts and Doppler reference frequencies either in 
the form of a constant frequency or uplink ramps. LaRa experiment will nominally 
perform tracking in the closed-loop mode. The LaRa transponder is designed to 
obtain closed-loop two-way Doppler measurements at the accuracy level \red{(see 
Table~\ref{tab:errbudg} for the details on the link budget)} of 0.05~mm/s at 60 
second integration time (or equivalently 2.8 mHz). Nevertheless, at long 
Earth-Mars distance, the signal-to-noise level at the ground stations will be so 
low that lock on the downlink carrier could be jeopardized and measurements 
acquisition might have to be performed in open-loop mode (this will be the case 
only if DSN 70~m antennas are not available, see the previous section). 
The open-loop system mixes an incoming intermediate frequency signal with a 
signal whose frequency is a linear approximation of the predicted frequency. The 
baseband signal is passed through a filter whose bandpass is centered at the 
expected frequency and has a sufficiently wide bandwidth to allow for any 
unexpected signal frequency excursions. A set of analog-to-digital converters 
digitizes the received bandwidth and then writes the samples onto open-loop data 
files. The (non-real time) processing of open-loop data basically consists of 
performing signal detection on the recorded samples using high accuracy signal 
parameter estimation algorithms. The received sky frequencies are then 
reconstructed from the detected frequencies. This method is used for very noisy 
signal for low SNR, but, as mentioned above, the limitation is not arising from 
the downlink SNR but rather from the uplink. It will not be necessary to use the 
open-loop method for the LaRa experiment.

\subsubsection{The error-budget}
\label{sec_errorbud}
The precision of the Doppler measurements for LaRa is limited by disturbing 
effects. Non-signal disturbances in a Doppler link are mainly due to 
\citep{Asmar:2005uq}:
\begin{itemize}
\item {The instrumental noises (random errors introduced by the ground station 
or LaRa),}
\item {The propagation noises (random frequency/phase fluctuations caused by 
refractive index fluctuations along the line-of-sight),}
\item {Systematic errors ({\it e.g.} ground station delay uncertainty).}
\end{itemize}
Instrumental noises include phase fluctuations associated with finite 
signal-to-noise ratio (SNR) on the radio links, ground and LaRa electronics 
noise, frequency standard noise, and antenna mechanical noise (unmodeled phase 
variation within the ground station) \citep[e.g.][]{Dehant:2009aa}. The larger 
contribution to the instrumental noise is the thermal noise that is related to 
the mean antenna system operating noise temperature. The propagation loss will 
vary as a function of the distance between Mars and the Earth. The thermal noise 
is of the order of $10^{-2}$~mm/s at 60-s for LaRa.
\\
The propagation noise \red{perturbs both the uplink and the downlink. It} is 
caused by phase scintillation as the deep space wave passes through media with 
random refractive index fluctuations such as solar plasma, Earth and Mars 
troposphere or ionosphere \citep{Asmar:2005uq,Zuber:2007fk,Bergeot+2019}.
The plasma Doppler shift is induced by all the charged particles along the radio 
wave path and depends on the elongation. 
Thus, the observations should be performed at times far from solar conjunction 
to avoid the solar plasma effects. A model adapted to Doppler data exists to 
estimate the standard deviation \red{(accounting for the two-way link)} due to 
solar phase scintillations in the range 5$^\circ$ $\le$ elongation angle $\le$ 
27$^\circ$(see \cite{Zuber:2007fk} and reference therein).\\
The Earth's ionosphere induces a frequency shift which can be partly removed by 
using models like, for example, the Klobuchar model \citep{Klobuchar:1987kx}.\\
Unlike the ionosphere, the troposphere (neutral atmosphere) is a non-dispersive 
medium, thus the propagation delay is not frequency-dependent. The total delay 
of radio signal caused by the neutral atmosphere depends on the refractivity 
along the traveled path, itself depending on pressure and temperature. There are 
two components: the dry component and the water vapor (the wet component). At 
X-band frequencies, tropospheric effects dominate the Doppler error with respect 
to ionospheric and solar plasma errors except for elongation smaller than 10-15º 
when the latter becomes dominant 
\citep[e.g.][]{Patzold:2016aa,Le-Maistre:2019ab}. The tropospheric delay due to 
dry air corresponds to 90$\%$ of the tropospheric effect. It depends mainly on 
well-known atmospheric pressure and temperature on the Earth’s surface and 
therefore it is easy to account for it from a model. The unmodeled remaining 
10$\%$ of total tropospheric delay due to the wet component depends on the water 
vapor content of the Earth's atmosphere in the propagation path, mainly below 
altitudes of 8–15 km \citep[e.g.][]{Zuber:2007fk}. This component is difficult 
to model without measurements due to the high spatial and temporal variability 
of the water vapor. However, the use of GNSS data or products will help to 
decrease the wet troposphere contamination of the observation 
\citep{Feltens:2018kx}. Finally, the ground station contributes to Doppler noise 
due to temperature and location uncertainties.\\
\red{The order of magnitude of Mars’ ionosphere effects have been determined by 
\citet{Bergeot+2019} from the subsurface mode of the Mars Express MARSIS radar. 
\citet{Bergeot+2019} developed also an empirical model of the Mars’ ionosphere 
called MoMo (MOdel of Mars iOnosphere). It is based on the large database of 
Total Electron Content (TEC) derived from the subsurface mode of MARSIS. The 
model provides the vertical TEC as a function of solar zenith angle, solar 
activity, solar longitude and the location. MoMo shows that Mars’ ionosphere 
variability is mainly driven by the solar illumination and activity, and the 
seasons, with amplitude variations of the vTEC over an entire day lower than ~2 
TECu. MoMo was used to estimate the impact on Doppler radioscience. At X-band, 
the maximum Doppler shift was estimated at the level of 0.05~mHz, or 0.001~mm/s, 
which is almost negligible but better to be estimated, in particular in view of 
the time variations of this correction.

The order of magnitude of the effects of the Martian troposphere on the Doppler 
signal has been estimated at a level near the Doppler instrument noise level. It 
can be estimated/calibrated using surface pressure and temperature measurements 
by the surface platform. This noise is well below the noise due to Earth water 
vapor fluctuations \citep{Folkner:2018aa}.}\\
All error contributions and their levels have been quantified, using the results 
from \cite{Zuber:2007fk} as reference. Table~\ref{tab:errbudg} shows the LaRa 
error budget \red{on the two-way Doppler}, assuming uncalibrated solar plasma 
and wet tropospheric effects as well as large ground station noise 
\red{(probably too large as the dynamics on the radio signal from an orbiter is 
larger than that on a lander)}. \red{Outside of the extreme cases (but still 
quite conservative), the RSS (Root Sum Square)} value reported in this table is 
about 0.07~mm/s@60s. \red{The RMS (Root Mean Square) value reported in this 
table is of the order of 0.03~mm/s~@60~s or maximum 0.05~mm/s~@60~s} (equal to 
2.8~mHz@60s in X-band) considered as representative noise level of LaRa 
measurements and used in the simulations shown below or in the companion papers 
\citep{Le-Maistre:2019ab,Peters:2019aa}.

\begin{table}[htbp]
\caption{Contributions to Doppler measurements errors for LaRa two-way X-band 
radio link integrated over 60 seconds.} 
\label{tab:errbudg}
\begin{center}
\footnotesize
\begin{tabularx}{\linewidth}{| X | X |}
\hline
\textbf{Error source} & \textbf{Two-way Doppler noise level}\\
                      & \textbf{(mm/s)} \\
\hline
Instrumental LaRa noise and thermal noise at ground station & $<$0.03 \\
\hline
Solar plasma effects at 10$^\circ$ and 25$^\circ$ elongation 
\citep{Zuber:2007fk} & 0.10 and 0.03\\
\hline
Earth Ionospheric effects (including scintillations) & 0.02 \\
Tropospheric effects of the Earth atmosphere wet component at 30º elevation 
angle (dry contribution is assumed properly corrected from pressure and 
temperature) \red{or dry atmosphere} & 0.06 \red{or $<$0.03}\\
\hline 
\red{Mars Ionospheric and tropospheric effects} & $<$0.03\\
\hline
Ground station & 0.04 \\
\hline
\red{Total Root Sum Square (RSS, computed as the square-root of the sum of the 
above square noise values)} & \red{0.13 and 0.07}\\
\hline
Total Root Mean Square \red{(RMS, computed as the square-root of  the mean of 
the sum of the above square noise values)} & \red{0.05} and \red{0.03}\\
\hline
\end{tabularx}
\end{center}
\end{table}

\subsubsection{MOP signatures in LaRa's measurements} 
\label{sec_MOPsign}
\begin{figure}
    \centering
    \includegraphics[scale=0.5]{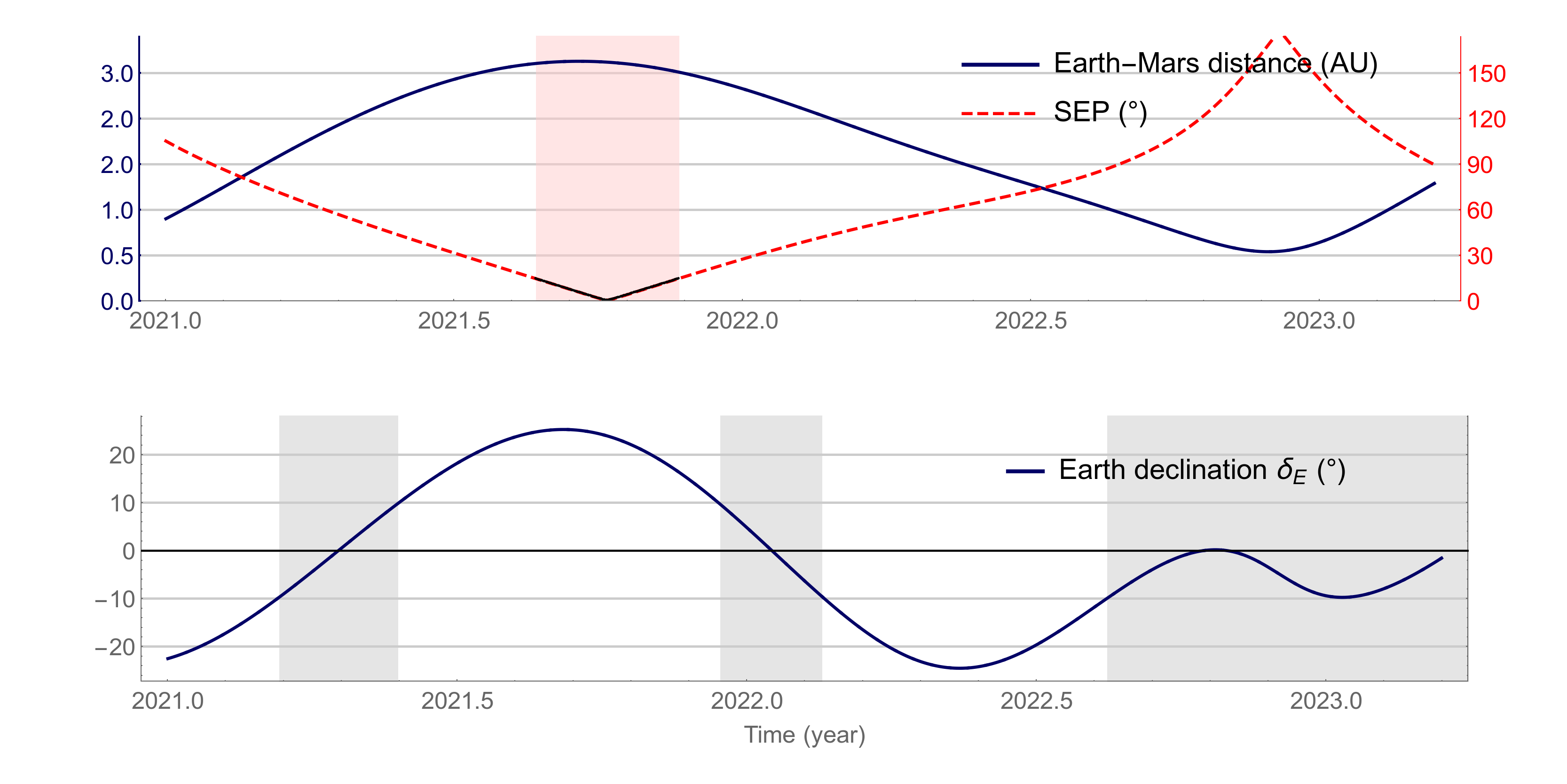}
    \caption{\red{Evolution of the Earth-Mars distance and the Earth declination 
as a function of the mission timing.}}
    \label{fig:geom}
\end{figure}

The sensitivity of the LaRa DTE Doppler measurements to the direction of the 
Martian spin axis, and hence to nutation is reduced when the Earth declination 
(the angle between the Mars-Earth vector and the Martian equatorial plane) is 
close to zero \citep{Yseboodt:2017if} because the Martian rotation axis is then 
perpendicular to the line-of-sight.
\red{The geometry of the mission and in particular, the evolution of the 
distance between Earth and Mars and the evolution of the Earth declination are 
shown in Fig.~\ref{fig:geom}.}

Since the Doppler observable is, to the first order, the projection of the 
lander velocity on the line-of-sight, a motion of the spin axis in space when 
the declination is close to zero results in a negligible contribution in the 
Doppler signal.
\\

\begin{table}[!htb]
\caption{Maximal value of the MOP signature in the Doppler observable (in mm/s) 
for the ExoMars 2020 mission. The nominal mission time interval is used. The 
lander position is in Oxia Planum (18.20$^\circ$~N latitude, 335.45$^\circ$~E 
longitude). 
See \citet{Folkner:2018aa}, Section 2, and Table \ref{tab:rotmod} for a 
description of the MOP model and the numerical values used.}
\centering
 $\begin{array}{|c|c|}
\hline
\text{MOP} &  \text{MOP signature (mm/s)} \\
\hline
\text{Nutations in obliquity} \quad \Delta\epsilon & 0.223 \\
\text{Nutations in longitude} \quad \Delta \psi & 0.253 \\
\text{Large liquid core ($T_{FCN} = -242$~d)} & 0.008 \\
\text{Small liquid core ($T_{FCN} = -285$~d)} & 0.004 \\
\text{Precession } (\Delta\dot\psi_0 = 2 \text{ mas/y}) & 0.051 \\
\text{LOD variations}  & 0.574 \\ 
\text{Polar motion} & 0.024 \\
\hline
\end{array}$
\label{tablsi}
\end{table}
The maximum values of the rotation angles signatures in the Doppler observable 
are given in Table \ref{tablsi}. The liquid core signature is computed assuming 
a Free Core Nutation (FCN) period of $-$242~days for a large liquid core or 
$-$285~days for a small liquid core. The largest signatures (up to 0.57~mm/s) 
are due to the Length-of-day (LOD) and the rigid nutation because they induce a 
large effect on the lander displacement (larger than 10 meters on Mars surface). 
The effects of the polar motion and of the liquid core in the Doppler signal are 
one or 2 orders of magnitude smaller (maximum 0.02~mm/s) for a lander with a 
$18^\circ$ latitude like ExoMars-2020.

\subsection{PRIDE measurements}
\label{pride}

Planetary Radio Interferometry and Doppler Experiment (PRIDE) has been developed 
at the Joint Institute for VLBI ERIC in collaboration with the Delft University 
of Technology and other partners. It is designed as an enhancement of planetary 
science missions' radioscience suite. The essence of PRIDE is in applying 
hardware and data handling algorithms developed for high-resolution radio 
astronomy studies of galactic and extragalactic objects using Very Long Baseline 
Interferometry (VLBI). In applications to interplanetary spacecraft as targets, 
VLBI observations must be treated in the so called near-field VLBI mode. This 
mode takes into account that the wave front of radio emission generated by 
spacecraft within Solar system arrives to Earth with non-negligible curvature 
(spherical in zero approximation) unlike the case of galactic and extragalactic 
sources, for which the wave front can be treated as planar. The main measurables 
of PRIDE are lateral coordinates of the target (radio emitting spacecraft) 
defined in the celestial reference frame (typically - the International 
Celestial Reference Frame, currently -- ICRF-3, as defined by the International 
Astronomical Union) and radial velocity of the target. The latter is deducted 
from Doppler measurements. 

The general methodology of PRIDE is described in \citep{Duev:2012uq} and 
specific details for Doppler extraction from PRIDE measurements in 
\citep{Bocanegra-Bahamon:2018aa}. To date, PRIDE was demonstrated in 
observations of the ESA's Mars Express \citep{Duev:2016aa,Molera+2017} and Venus 
Express \citep{TMBB+2019}. As demonstrated in multiple observations of Mars 
Express, PRIDE can achieve Doppler measurements in a three-way regime with the 
median noise level of $\sim 2$~mHz (0.03~mm/s) in 10~s integration or $\sim 
0.013$~mm/s in 60~s integration \citep{Duev:2016aa}. PRIDE is adopted as one of 
eleven experiments of the ESA's flagship mission JUICE (Jupiter Icy moons 
Explorer) scheduled for launch in 2022 \citep{JUICE-2015}. 

\red{In the LaRa context, we intend to use PRIDE as a supplementary technique, 
by no means replacing the nominal two-way radio measurements conducted by DSN 
facilities. However, a}s demonstrated recently in observations of {\it e.g.}, 
ESA's Venus Express, due to a wider choice of ground-based receiving stations 
among \red{PRIDE} radio telescopes, some of them, at a particular time, might be 
in a more favorable position (a higher elevation of a target, a better seasonal 
climatic conditions, a less turbulent patch of ionosphere over the ground 
antenna, etc.) for achieving Doppler measurements with a very low noise (see, 
{\it e.g.}, Fig.~6 in \citep{TMBB+2019})\red{, lower than a single nominal 
dedicated DSN or ESTRACK facility. This simple PRIDE advantage might help to 
achieve the science goals of LaRa}.

The Doppler measurements extracted from PRIDE observations contain the same 
information of interest as contained in the measurements recorded at DSN but 
they are statistically independent from those nominal measurements in noise 
contributions defined by the downlink propagation and ground receiving 
instrumentation. \red{Thus, we intend to conduct PRIDE measurements in parallel 
with nominal two-way LaRa radio sessions. A detailed error propagation analysis 
in PRIDE measurements is given in Section~3.3 of \citet{TMBB+2019}. We note that 
the overall setup of the nominal two-way configuration of LaRa experiment is 
essentially the same as in Venus Express observations described by 
\citet{TMBB+2019}, thus making the mentioned error propagation directly 
applicable to the LaRa case. However, in the months prior to LaRa operations we 
intend to conduct special verification PRIDE observations of operational Mars 
landers and orbiters in order to tune up observational setup and data processing 
algorithms.}

\red{In addition, PRIDE might be exercised in one-way (downlink only) radio 
sessions.} The contribution of PRIDE measurements in \red{this} non-coherent 
(one-way) regime is at least in providing test and calibration information. This 
one-way regime will be investigated further in the months preceding the ExoMars 
mission launch.

\subsection{LaRa operations}
After landing, the measurements will be carried out twice per week during the 
whole mission lifetime (twice per week during the minimum guaranteed mission of 
one Earth year and during the extended mission, with a possible relaxation to 
once per week during lander hibernation seasons if any). No operation will be 
done at solar conjunction and for a solar elongation angle lower than~5$^\circ$ 
\red{and special care (such as avoiding solar irruption period) will have to be 
performed at solar elongation angle between 5$^\circ$ and 10$^\circ$}. Each 
tracking session will last one hour and will be performed during the local day 
time at Mars because of power-availability on the lander. LaRa is designed to 
allow communication with the Earth when the line-of-sight to the ground station 
is between $30^\circ-55^\circ$ of elevation. Nevertheless, we plan to operate 
LaRa preferably when the Earth appears between 35$^\circ$ and 45$^\circ$ of 
elevation above the lander's horizon because (1) this is the elevation range of 
LaRa antennas best gain (see Fig.~\ref{fig3Dpattern}), (2) this limits Mars 
atmospheric effects and problems related to local reflections of the signal and 
local topography blocking the view to the Earth, (3) this is the elevation range 
of best sensitivity to the nutation parameters \citep{Le-Maistre:2019ab}. First 
tracking session of each week will be performed from lander's east direction 
(Earth rise) and second one from lander's west direction (Earth set), or {\it 
vice versa}, the first tracking session of each week can be performed from 
lander's west direction (Earth set) and second one from lander's east direction 
(Earth rise) as it is important to alternate the line-of-sight azimuth direction 
from one pass to the other \citep{Le-Maistre:2019ab}. Earth tracking stations to 
be used will be preferably those that have low water content in the atmosphere, 
reducing Earth tropospheric disturbances. Additional constraints on LaRa 
operations from other units of the SP (system and scientific payload) are 
expected to avoid inter-units interference. For instance, LaRa will certainly 
not operate simultaneously with the Ultra High Frequency (UHF) transponder used 
for Telemetry and Telecommand (TM/TC) communications with Martian orbiters. LaRa 
will also have to wait for ExoMars rover egress before starting to operate for 
possible interference reasons.

\section{LaRa expected results}
\label{sectionMOP}
\subsection{MOP determination}
We carry out numerical simulations with the GINS (G\'eod\'esie par 
Int\'egrations Num\'eriques Simultan\'ees) software in order to assess the 
uncertainties with which LaRa will allow to determine the Mars Orientation and 
rotation Parameters (MOP) as well as the lander coordinates at the surface of 
Mars. The actual rotation model used to create the synthetic Doppler data is 
classical and described in details in \cite{Le-Maistre:2018ab}. Basically, the 
angles at epoch ($\varepsilon_0,\psi_0,\phi_0$), rates ($\dot 
\varepsilon_0,\dot\psi_0, \dot\phi_0$) and spin angle variation amplitudes 
($\phi_m^{c/s}$) are from \cite{Konopliv:2016aa}. The amplitudes of the truth 
model of nutation are computed from the model of rigid nutation of 
\cite{Roosbeek:1999lh} or \cite{Baland:2019in}, modulated by a liquid core of 
FCN period equal to $-240$ days and of amplification factor $F$ equal to 0.07. 
The CW amplitudes are arbitrarily chosen and its period is fixed (and not 
estimated) to be around 200 days for the non-hydrostatic case and 220 days for 
the hydrostatic case. The numerical values of the MOP that are estimated are 
reported in Tab.~\ref{tab:rotmod} along with the post-fit uncertainties for 
seven different scenarios:
\begin{itemize}
\item [sol1:] This is the solution obtained after one Earth year of data 
accumulation (ExoMars 2020 nominal mission). The nutation parameters estimated 
here are the amplitudes of the complex prograde ($p_m$) and retrograde ($r_m$) 
nutation terms at the forcing period $m=2$ and $m=3$, corresponding to the 
semi-annual period of 343.5 days and the ter-annual period of 229 days, 
respectively. Obliquity rate ($\dot \varepsilon_0$) and precession rate 
($\dot\psi_0$) of the spin axis at epoch J2000 are estimated as well as the 
annual ($\phi_1^{c/s}$) and semi-annual ($\phi_2^{c/s}$) amplitudes of the Mars 
spin angle. See details on this setting in \cite{Le-Maistre:2018ab}.
\item [sol2:] This solution has the same settings as sol1 parameter-wise, but is 
computed over one Martian year ($\sim$700 days corresponding to LaRa nominal 
lifetime).
\item [sol3:] This solution is similar to sol2, but the annual amplitudes of 
nutation ($p_1,r_1$) are also estimated.
\item [sol4:] This solution is similar to sol2, but the spin angle amplitudes of 
the ter-annual ($\phi_3^{c/s}$) and quater-annual ($\phi_4^{c/s}$) periods are 
also estimated.
\item [sol5] This solution is similar to sol2, but the nutation parameters 
estimated are the parameters of the nutation transfer function: $F$ and 
$T_{FCN}$ (and no more the amplitudes of the nutation series).
\item [sol6] This solution is similar to sol2, but the amplitudes of the CW are 
also estimated.
\item [sol7] This solution is similar to sol3 (parameter-wise), but is obtained 
by combining one Mars year of RISE (InSight) Doppler data together with one Mars 
year of LaRa Doppler data. See details on this scenario in 
\cite{Peters:2019aa}. 
\end{itemize}
Generally speaking, we see from Tab.~\ref{tab:rotmod} that LaRa will allow us to 
determine  the nutation signal very precisely, with only a few mas of 
uncertainty on the amplitudes and about 4 days on the FCN period. If the latter 
estimation is actually depending on the parameter setup, i.e.~the a priori value 
taken for $T_{FCN}$ (here $-$240 days), the former is not. One can thus 
confidently claim that the two most-interesting parameters, $p_2$ and $r_3$, 
will be estimated by LaRa alone with 4-7~mas and 4-8~mas of uncertainty, 
respectively (depending on the set of estimated parameters, and for the LaRa 
nominal lifetime). One could even reduce those uncertainty ranges by about 2~mas 
if we use RISE and LaRa data together. Such an apparently small difference in 
the nutation estimate uncertainties can significantly enhance the scientific 
return of those two missions as shown in the following section and further 
discussed in \cite{Peters:2019aa}. Spin angle variations and precession rate 
($−7608.3 \pm 2.1$~mas/year) are already accurately known thanks to historical 
lander, rover and orbiter data \citep{Kuchynka:2013uq,Konopliv:2016aa}, but LaRa 
shall definitely further improve the determination of those parameters. The 
precession uncertainty shall decrease from 2.1 to 0.3~mas/year. This will allow 
us to get one order of magnitude better on the moment of inertia of the whole 
planet (presently at the level of $0.3638 \pm 0.0001$), which is one of the main 
parameters that constrain the interior models \citep{Rivoldini:2011aa}. One 
major discovery of LaRa could be the first-time direct measurement of the 
Chandler wobble amplitudes as shown in sol6. Forced polar motion in the Earth 
system is superposed by free oscillations of the Earth, i.e.~CW.  As the CW is a 
resonance oscillation of the planet, potential excitation mechanisms require 
energy in a band close to the CW frequency in order to sustain the free polar 
motion and thus to counteract its damping. On the Earth the CW is excited by a 
combined effect of atmosphere and ocean.  If this free mode  exist on Mars, it 
would provide novel information on Mars' atmospheric dynamics  in a band close 
to the CW frequency. Because of their low latitudes, none of the previous landed 
missions have been able to detect this signal, neither will the near-equatorial 
InSight mission. The  
detection of the CW frequency with LaRa will put some constraints on the 
interior of Mars, while the estimation of the amplitude will put constraints on 
the dissipation inside Mars, as well as on the energy in the atmosphere near the 
CW frequency.\\
\\
Finally, because of the high accuracy of LaRa and RISE forthcoming data, one 
could expect that the tidal Love number $h_2$, characterizing the radial 
deformation of Mars in response to tidal forcing, could be measured for the 
first time. Additional simulations (not shown in Tab.~\ref{tab:rotmod}) show 
that $h_2$ can be estimated from LaRa alone with an uncertainty of 0.6 while a 
combination of RISE and LaRa data would allow estimating $h_2$ with an 
uncertainty of 0.2. This is respectively three times and one time the expected 
value for $h_2$ \citep{Van-Hoolst:2003uq}, which makes the determination of this 
parameter with RISE or LaRa not relevant. \\
\\
\begin{table}[htbp]
\caption{Mars rotation model to retrieve, along with the a priori constraints 
and post-fit uncertainties in the MOP estimates obtained with GINS using an 
iterative least square procedure. The nutation amplitudes are given for the 
rigid case \citep[see][]{Baland:2019in} and for the fluid core case 
(rigid/fluid), considering the nominal values for the Core momentum factor 
(liquid core amplification factor in Section 2 after Eq.~(\ref{eq:TF})) and the 
FCN period. \red{The LOD amplitudes are taken from \citet{Konopliv:2016aa}.}} 
 \label{tab:rotmod}
 \begin{center}
 \footnotesize
 \begin{tabular}{ l l l l l l l l l l}
  Parameter&Symbol&Nominal&sol1&sol2&sol3&sol4&sol5&sol6&sol7 \\
            &      &(Truth) &nom.&ext.&nut1&lod4&FCN &CW &RISE \\
Obliquity rate (mas/year)&$\dot \varepsilon_0$&-2.0&0.7 &0.2&0.2&0.4&0.1&0.2&0.1 
\\
Precession rate (mas/year)&$\dot \psi_0$&-7608.3&1.5&0.4&0.5 &0.9&0.3&0.6&0.3\\ 
1-annual cosine spin angle (mas)&$\phi^c_1$&481&17&1&8&2&2&2&5 \\
1/2-annual cosine spin angle (mas)&$\phi^c_2$&-103&16& 6&8&9&3&6&5 \\
1/3-annual cosine spin angle (mas)&$\phi^c_3$&-35&&&&7 \\
1/4-annual cosine spin angle (mas)&$\phi^c_4$&-10&&&&1 \\

1-annual sine spin angle (mas)&$\phi^s_1$&-155&18&2&7&4&3&3&6\\ 
1/2-annual sine spin angle (mas)&$\phi^s_2$&-93&20&6&10&7&5&8&5\\ 
1/3-annual sine spin angle (mas)&$\phi^s_3$&-3&&&&11 \\
1/4-annual sine spin angle (mas)&$\phi^s_4$&-8&&&&1 \\
Core momentum factor (a priori)&$F$&0.07&&&&&0.02 \\
FCN period (days)&$ T_{FCN}$&-240&&&&&4 \\
1-annual prograde nutation (mas)&$p_1$&102/104&&&5&&&&4 \\
1/2-annual prograde nutation (mas)&$p_2$&498/512&11&4&6&7&&5&4 \\ 
1/3-annual prograde nutation (mas)&$p_3$&108/112& 5&3&3&4&&4&2 \\

1-annual retrograde nutation (mas)& $r_1$&137/132&&&6&&&&4 \\
1/2-annual retrograde nutation (mas)&$r_2$&18/15& 15&4&8&5&&5&3\\ 
1/3-annual retrograde nutation (mas)&$r_3$&5/12&10&4&4&8&&6&2 \\
Chandler wobble cosine (mas)&$X^c_{cw}$&33&&&&&&4 \\
Chandler wobble sine (mas) &$X^s_{cw}$&33&&&&&&4 \\
Chandler wobble cosine (mas)&$Y^c_{cw}$&33&&&&&&6 \\
Chandler wobble sine (mas)&$Y^s_{cw}$&33&&&&&&4 \\
  \end{tabular}
 \end{center}
\end{table}

\subsection{Mars interior structure and atmosphere}  

\subsubsection{Constraints on the interior of Mars}
In order to illustrate and quantify how the interior structure of Mars affects 
its wobble and nutation we calculate the FCN period and the amplification 
effects on the largest nutations (retrograde annual, semi-annual, and ter-annual 
and prograde semi-annual) for a large set of plausible Mars interior structure 
models.\\
\\
These interior models are built in two steps. First, we assume that the planet 
is spherical, isotropic, elastic, non-rotating, and in hydrostatic equilibrium. 
The planet has a crust that is modeled by its thickness and density, a silicate 
mantle, and a fluid and iron-sulfur core. The range of densities and thicknesses 
of the crust are in agreement with \citet{Wieczorek:2004nl}. For the mantle, we 
consider five plausible Mars mantle compositions 
\citep{Taylor:2013fk,Sanloup:1999gj,Lodders:1997rv,Mohapatra:2003zc,
Morgan:1979mm} (denoted in the figures as: DW, EH45, LF, MM, and MA; these 
couples of letters usually refer to the names of the authors of the papers on 
mantle compositions, except for EH45) and two temperature end-members that have 
been deduced from 3D thermal-evolution studies \citep{Plesa:2016aa}. The 
compositions have been deduced from Martian meteorites and assumptions about 
Mars' formation. Elastic properties of the mantle are calculated with {\tt 
PerpleX} \citep{Connolly:2005aa} and those of the liquid Fe-S core are detailed 
in \citet{Rivoldini:2013uq}.
 In order to calculate the interior structure, we integrate the Poisson equation 
(relating the gravitational potential to the density) and hydrostatic pressure 
equation (relating the pressure exerted at equilibrium to the gravity and 
density) in the whole planet and the adiabatic gradient equation in the core. 
Our models match Mars' mass exactly, assume zero pressure at the surface, and 
continuity of pressure and gravity between crust, mantle, and core and 
continuity in temperature between the lower mantle and core. The system of 
ordinary differential equations and the associated boundary values are solved 
with {\tt BVP Solver} (Boundary Value Problem) \citep{Shampine:2006aa}. We have 
only retained models that agree at $1\sigma$ with the average moment of inertia 
($\mathrm{MoI}=0.3638 \pm 0.0001$ \citep{Konopliv:2016aa}). Models with a solid 
inner core can also agree with the $\mathrm{MoI}$ but have denser and smaller 
cores and therefore do not agree with the tidal Love number $k_2$ 
\citep{Rivoldini:2011fk}. For this reason, we do not consider models with a 
solid inner core in this study.\\

In a second step, the spherical models are deformed hydrostatically, i.e.~it is 
assumed that they are fluid and rotating with the rotation period of Mars. The 
geometric flattening of the hydrostatic models are calculated by solving 
Clairaut's equation \citep[e.g.][]{Moritz:1990fk}. The polar and equatorial 
moments of inertia of the planet and core can \red{then be directly} calculated 
\red{starting from} the geometric flattening.\\

\begin{figure}[htbp]
	\centering
		\includegraphics[height=7cm]{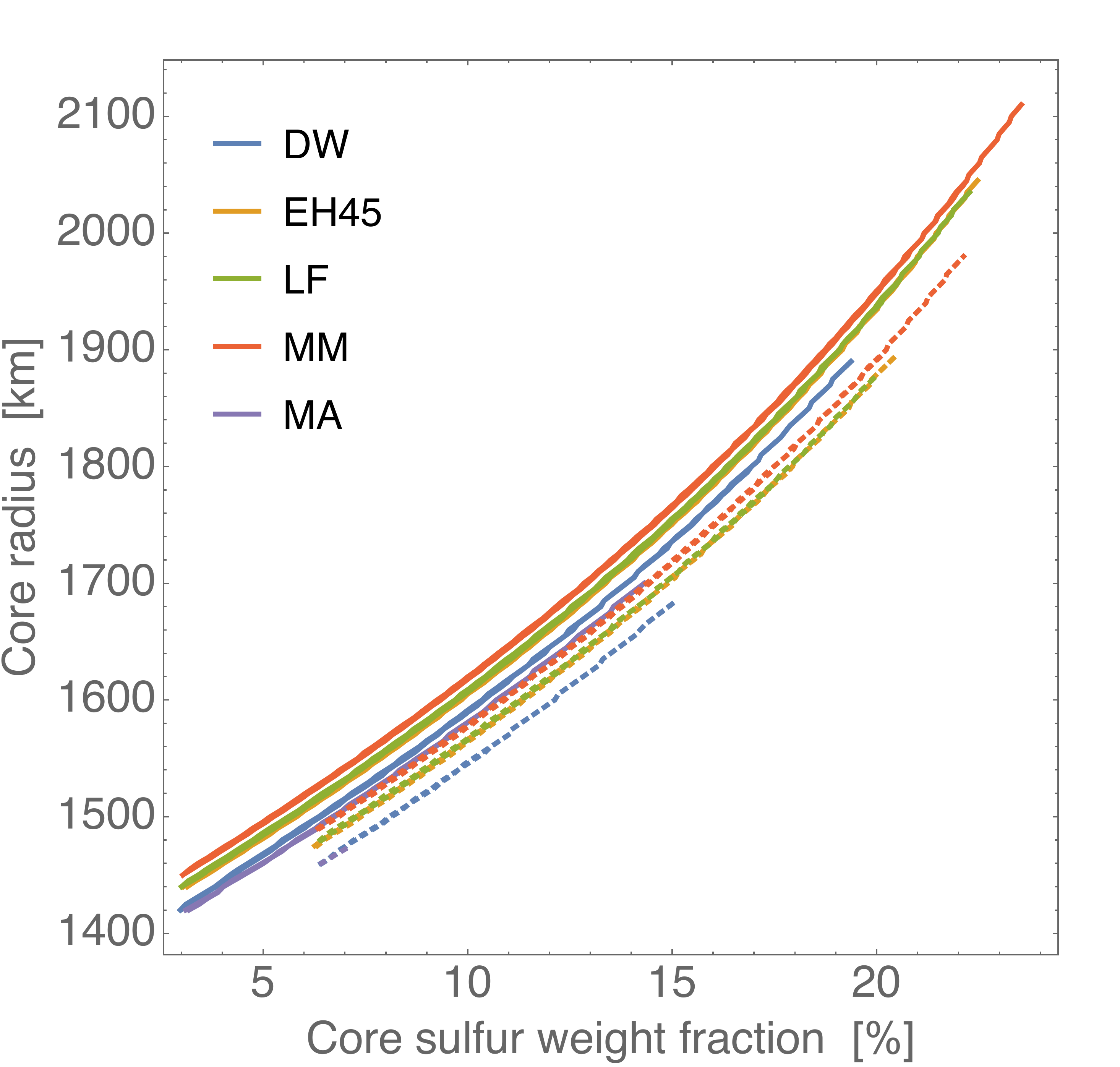}
	\caption{Core radius as a function of core sulfur concentration. Models 
using the cold temperature profile are shown by dotted lines.}
	\label{fig:rcmb(xS)}
\end{figure}

The relation between core sulfur concentration and radius for the different 
mantle models and temperature profiles is shown in Fig.~\ref{fig:rcmb(xS)}. The 
core radius range of the models results from the requirement that the models 
have a fully molten core (lower boundary) and are within $1\sigma$ of the 
$\mathrm{MoI}$ of Mars. Unlike models with the hotter mantle temperature, cold 
mantle models require more sulfur to remain completely molten and have a denser 
mantle and, therefore, a somewhat larger mantle moment of inertia at a given 
core radius. For this reason, the range of core radii that agrees with the 
$\mathrm{MoI}$ is reduced compared to hot mantle models.\\

\begin{figure}[htbp]
	\centering
\subfloat[]{\includegraphics[height=7cm]{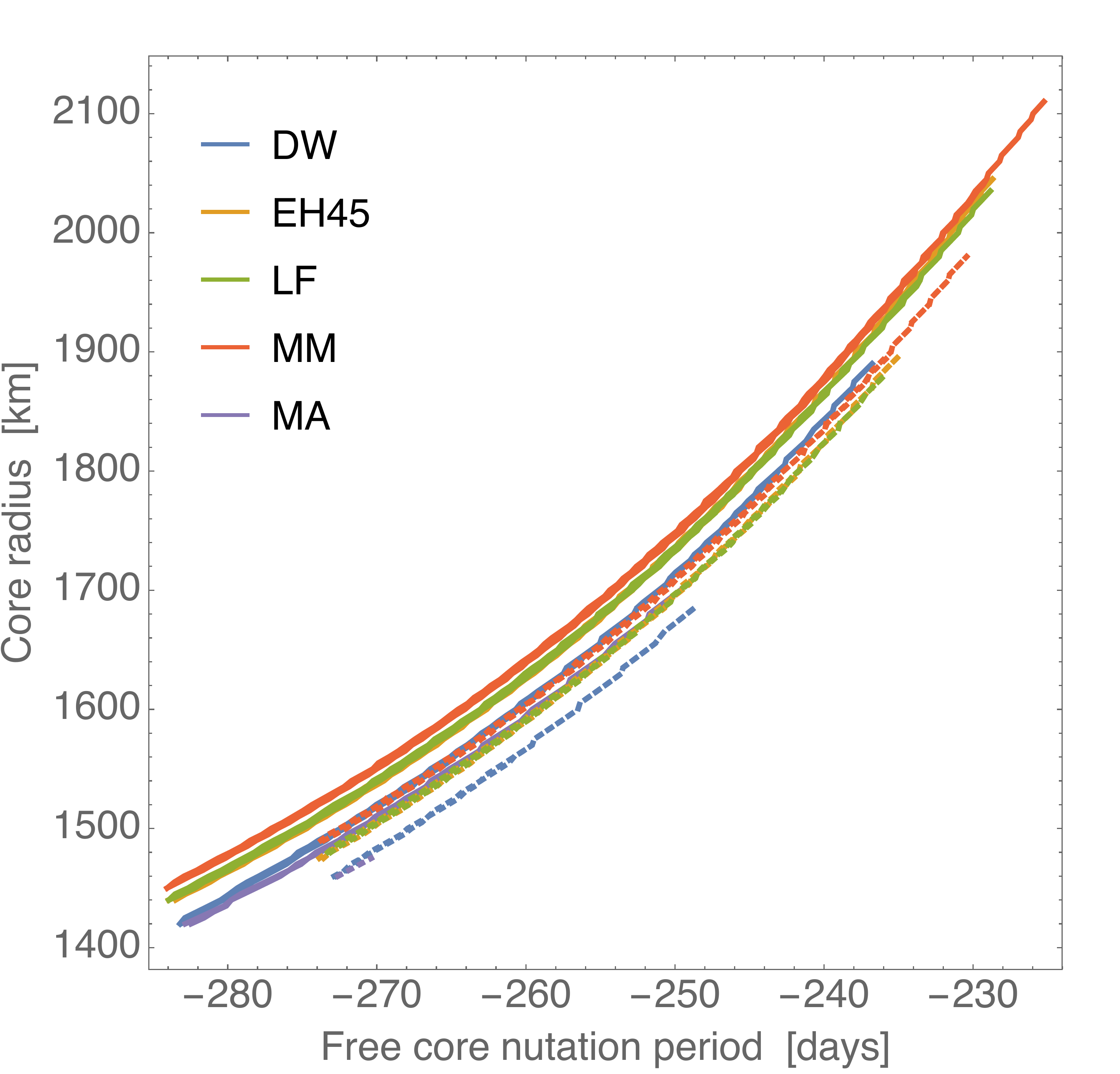} 
\label{fig:rcmb(fcn)}}
\subfloat[]{\includegraphics[height=7cm]{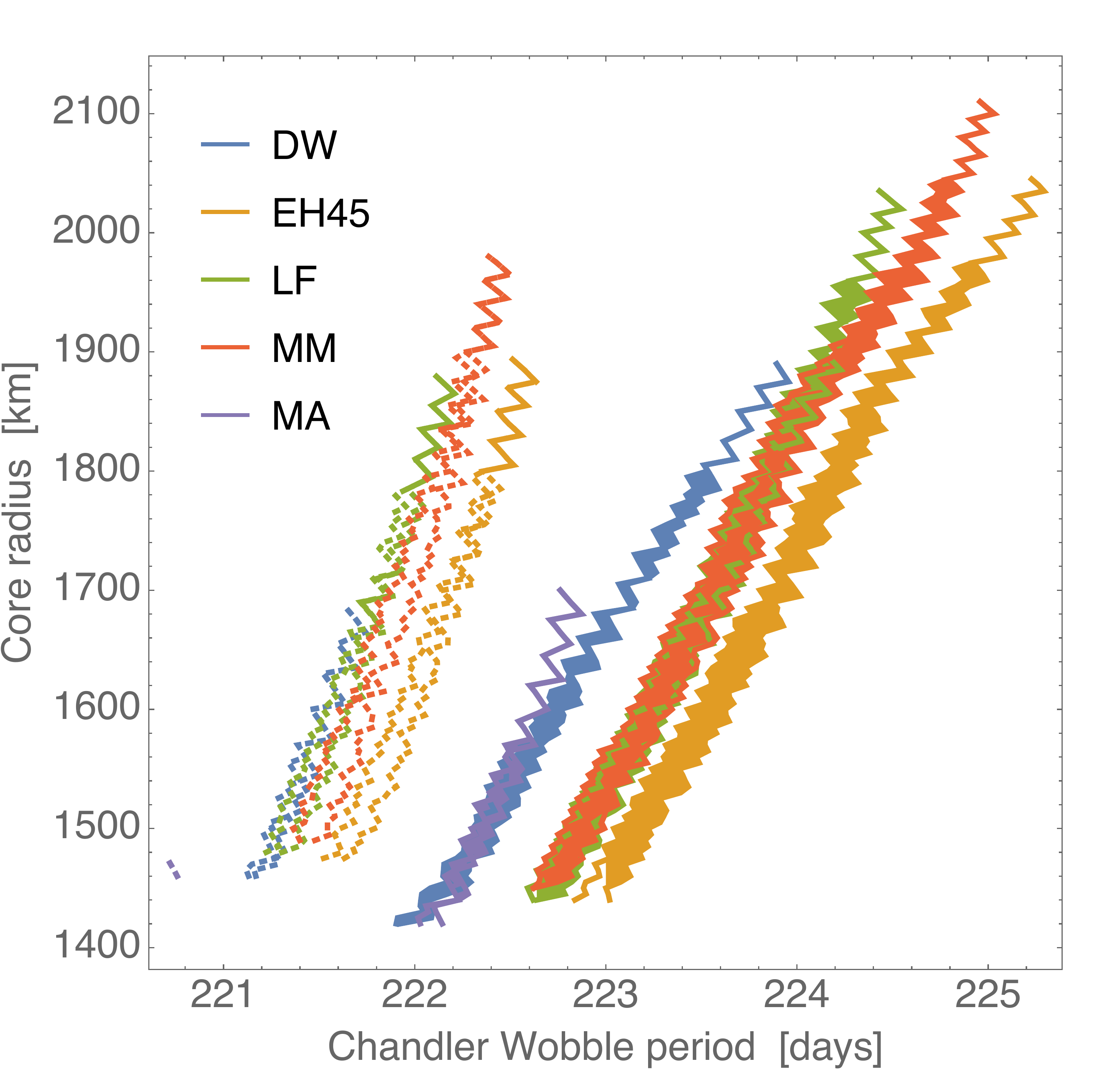} \label{fig:rcmb(CW)}}
\caption{Core radius as a function of \protect\subref{fig:rcmb(fcn)} 
Free Core Nutation period (in Inertial Frame) and \protect\subref{fig:rcmb(CW)} 
Chandler Wobble period (in co-rotating Body Frame). The models are built on an 
initial state in hydrostatic equilibrium. Models using the cold temperature 
profile are shown by dotted lines.}
\end{figure}

The period of the FCN (Fig.~\ref{fig:rcmb(fcn)}) increases with core radius 
because the dynamical flattening of the core mantle boundary increases faster 
than the moment inertia of the mantle (Eq.~(\ref{eq:FCN2axial})). The FCN 
periods of the models are within the periods of the retrograde semi-annual and 
ter-annual ($-$343.5~days to $-$229~days) nutations and it can be expected that 
these nutations are the most affected by the liquid core (changes of $16\%$ and 
$140\%$ for $r_2$ and $r_3$, respectively, for the truth model of 
Tab.~\ref{tab:rotmod}).\\

The period of the CW (Fig.~\ref{fig:rcmb(CW)}) increases with core radius 
because the compliance $\kappa$ increases faster than the moment of inertia of 
the mantle decreases (Eq.~(\ref{eq:CW})). For all the considered models, the 
variation is less than 5~days, smaller than the expected precision of LaRa. This 
is the reason why this parameter is not estimated but fixed using its 
theoretical value. \\

\begin{figure}[htbp]
\centering
\subfloat[]{\includegraphics[height=7cm]{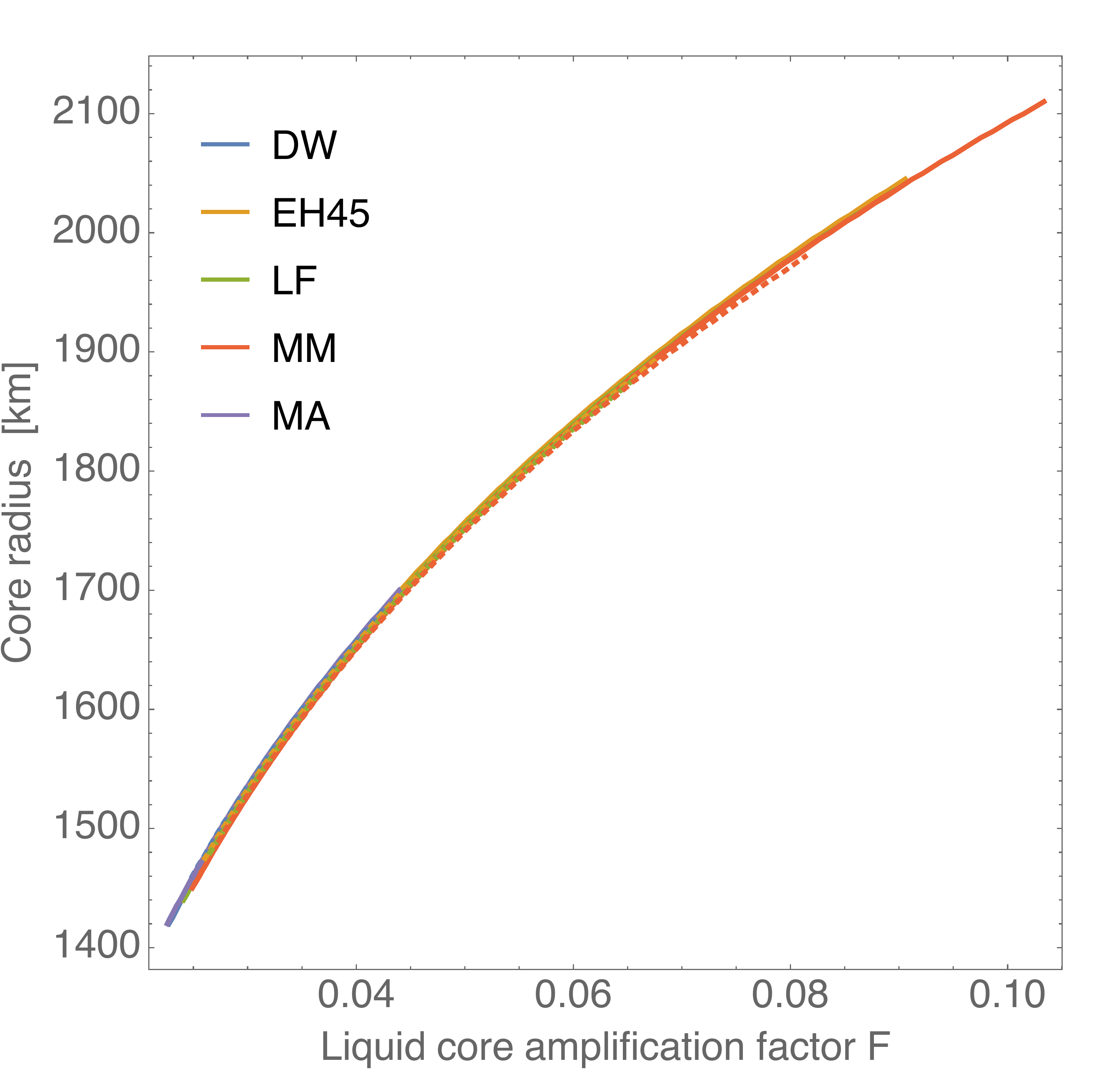} \label{fig:rcmb(F)}}
\subfloat[]{\includegraphics[height=7cm]{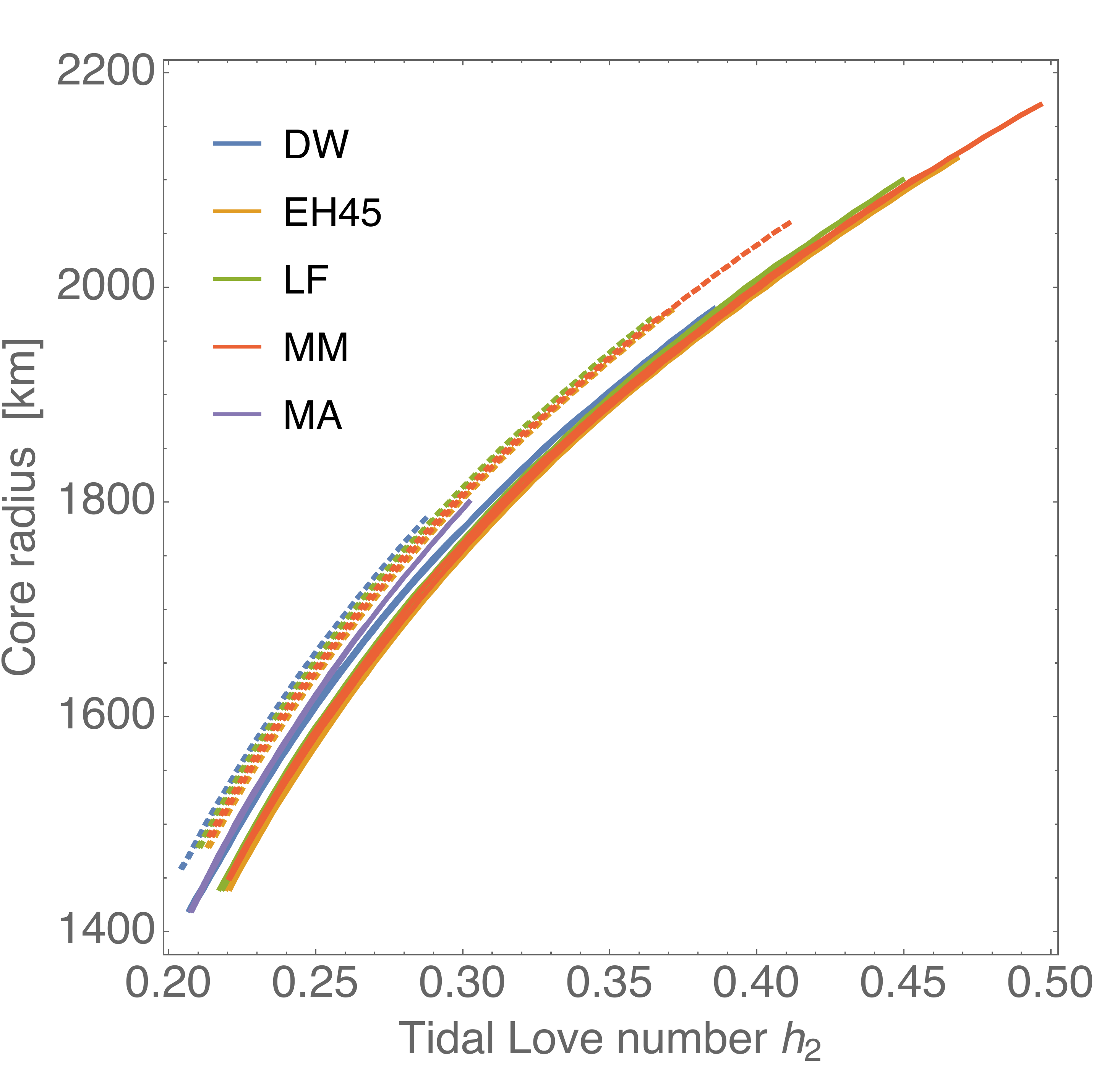} \label{fig:rcmb(h2)}}
\caption{Core radius as a function of \protect\subref{fig:rcmb(F)} 
liquid core amplification factor $F$ (in Inertial Frame) and 
\protect\subref{fig:rcmb(h2)} tidal Love number $h_2$ (in co-rotating Body 
Frame). Models using the cold temperature profile are shown by dotted lines.}
\end{figure}

The liquid core amplification factor $F$ (Fig.~\ref{fig:rcmb(F)}) increases with 
core radius or increasing moment of inertia of the core (Eq.~(\ref{eq:TF})). It 
is only weakly dependent on the thermal state of the planet and mineralogy of 
the mantle and is therefore a robust estimator for the core radius. With the 
expected precision of LaRa \red{and in the situation that the FCN resonance is 
weak}, the core radius can be estimated with a precision of about 160 km 
assuming we will get a precision of 0.02 on $F$.\\

\begin{figure}[htbp]
\centering
\subfloat[]{\includegraphics[height=6.5cm]{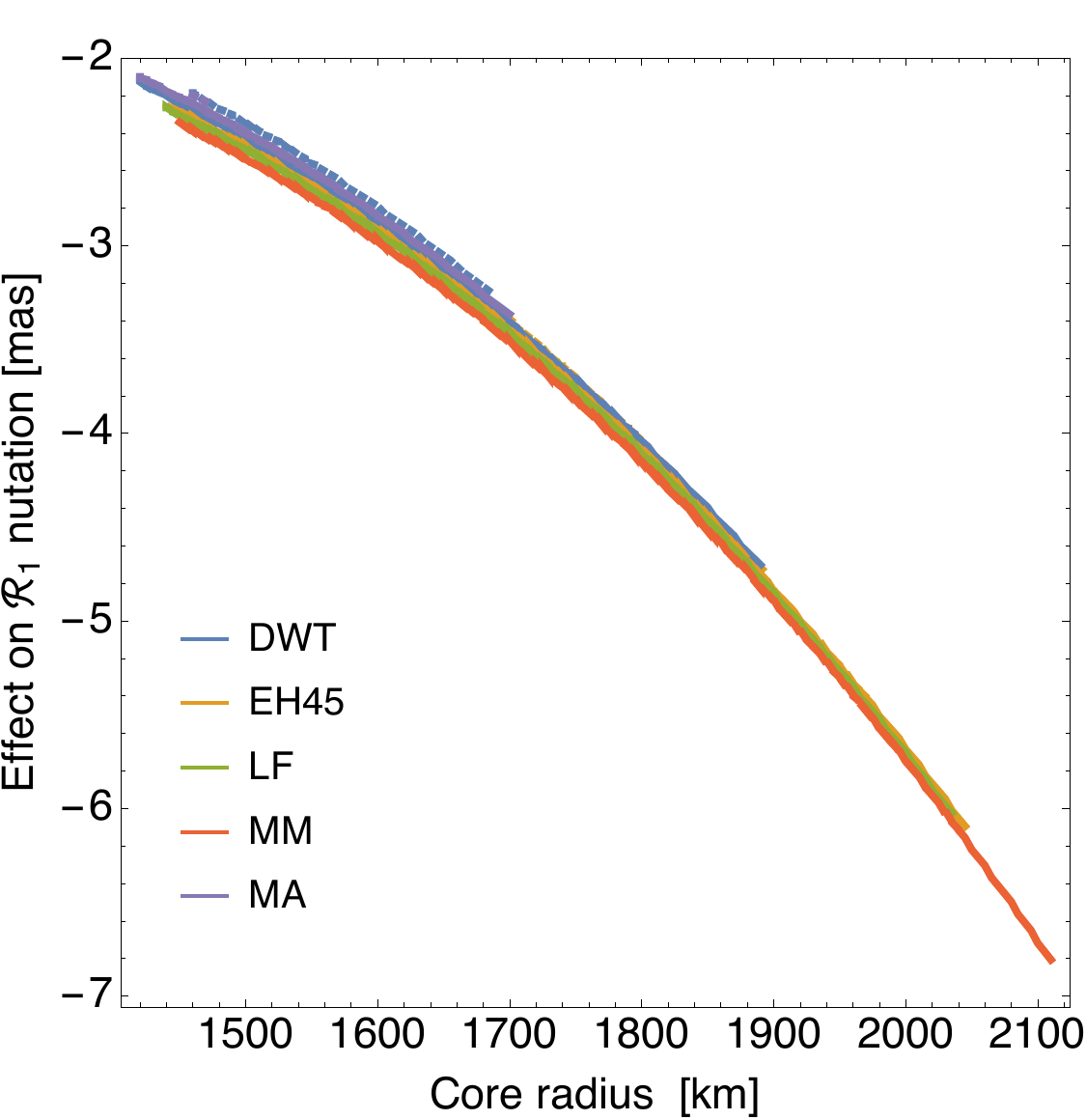}
\label{fig:retroAnnual}} \hspace{0.5cm}
\subfloat[]{\includegraphics[height=6.5cm]{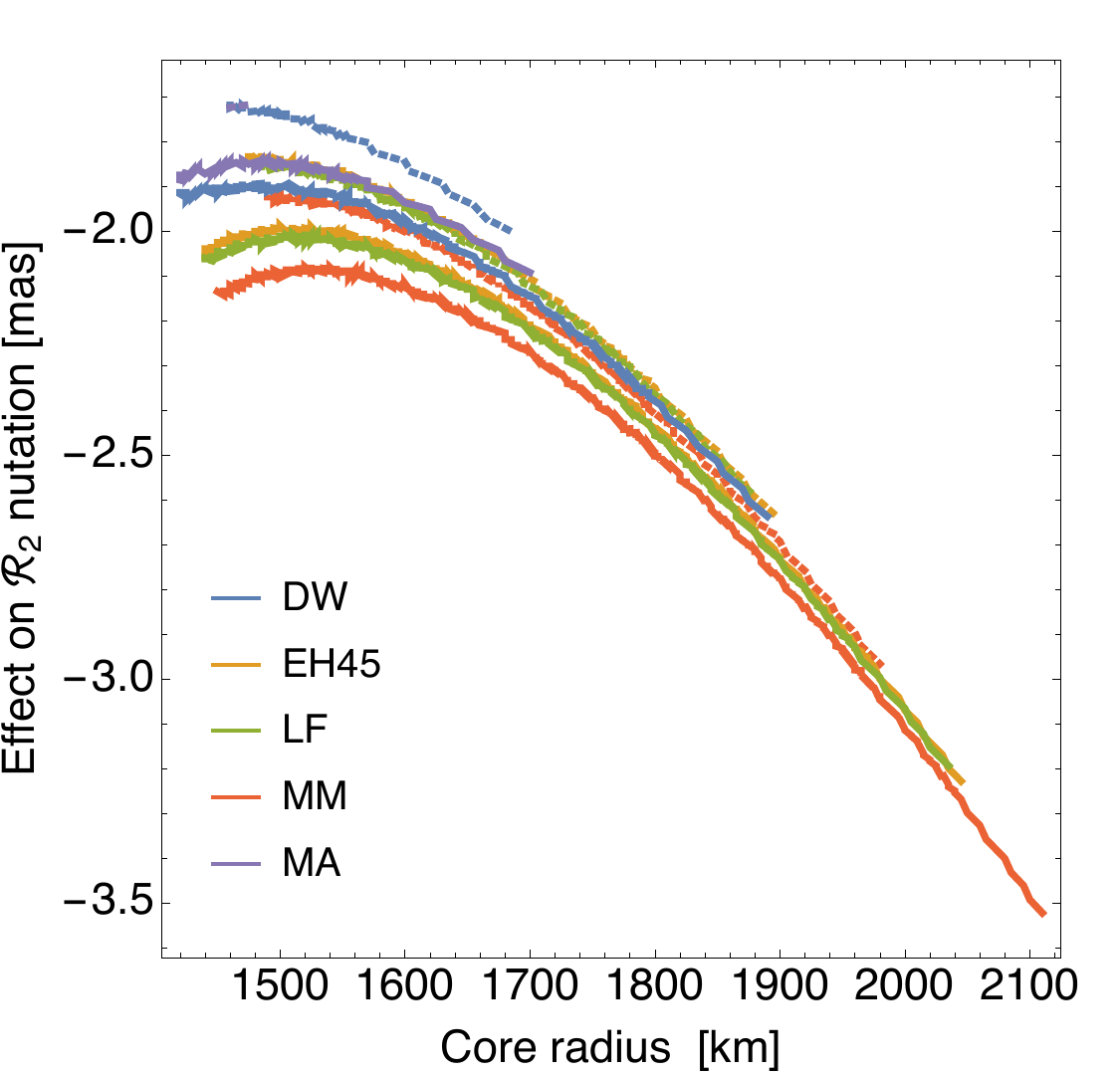}
\label{fig:retroSemiAnnual}}\\ 
\subfloat[]{\includegraphics[height=6.5cm]{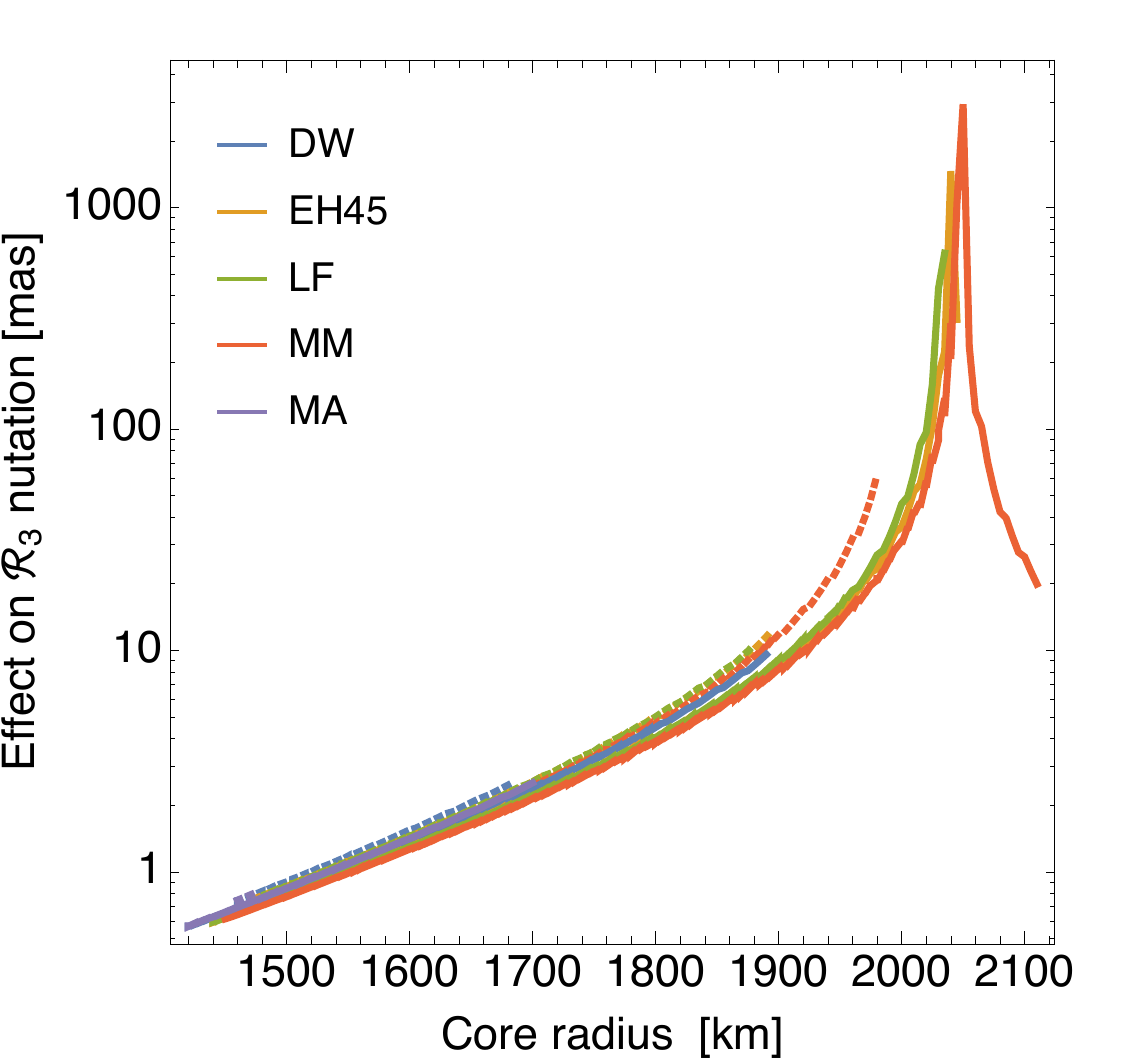}
\label{fig:retroTerAnnual} }\hspace{0.5cm}
\subfloat[]{\includegraphics[height=6.5cm]{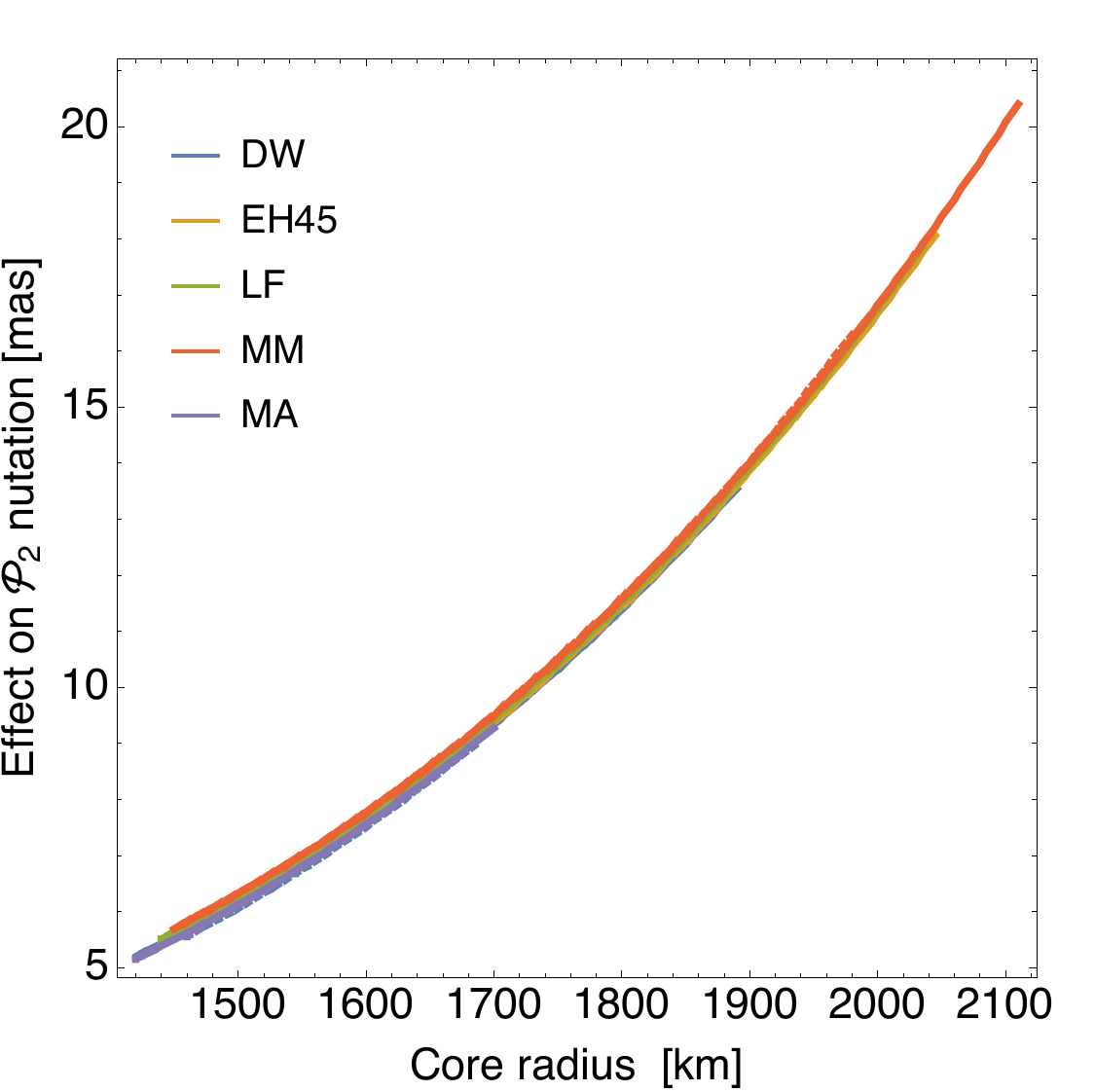}
\label{fig:proSemiAnnual}}
\caption{Effect of FCN rotational normal mode on 
(a) retrograde annual, 
(b) semi-annual, 
(c) ter-annual, and 
(d) prograde semi-annual nutations as a function 
of core radius. Models using the cold temperature profile are shown by dotted lines.}
	\label{fig:AmplifiedNuatations}
\end{figure}

Among the considered nutations, the retrograde ter-annual rigid nutations is the 
most affected by the FCN normal mode (Fig.~\ref{fig:AmplifiedNuatations}). In 
the hydrostatic equilibrium case, models with core radii about 2050~km could 
experience amplifications in excess of 100~mas because they have an FCN period 
that is very close to that of the retrograde ter-annual rigid nutation. Models 
with such large cores do, however, not agree with the observed tidal $k_2$ Love 
number at $1\sigma$ 
\citep{Genova:2016aa,Konopliv:2016aa,Rivoldini:2011fk,Khan:2018aa}. Geodesy 
constraints require that the core radius of Mars is between 1730~km and 1859~km 
\citep{Rivoldini:2011fk,Khan:2018aa} (computed for hydrostatic equilibrium). For 
the whole set of core radii, a variation of the $r_1$ amplitude by 2~to 7~mas 
could be expected whereas $r_3$ can be amplified by up to few 100~mas and $p_2$ 
can vary between 5~and~20~mas (Fig.~\ref{fig:AmplifiedNuatations}).
With the expected precision provided by LaRa alone of 4~to~8~mas in $r_3$ and 
4-7~mas in $p_2$ (see Tab.~\ref{tab:rotmod}), the radius of the core could then 
be estimated with a precision of 1~to~300~km depending on the radius of the 
core. 
When using both RISE and LaRa (precision of 2~mas and 4~mas on $r_3$ and $p_2$), 
this estimation can be reduced and the precision ranges from 1~to~100~km 
depending on the radius of the core.

It is well known that the shape of Mars deviates significantly from the 
ellipsoidal shape of a purely hydrostatic planet of the same composition and 
size as Mars deformed from a fluid sphere by rotation. It is thought that this 
deviation is mainly the result of the large Martian volcanic provinces 
\citep[e.g.][]{Phillips:2001aa}. It can be expected that these mass anomalies 
also affect the hydrostatic shape of the core  through their loading effect. 
Since the frequency of the FCN is almost linearly dependent on the dynamic 
flattening of the core (Eq.~(\ref{eq:FCN2axial})), core radius estimations from 
the FCN period could be biased if the correct shape of the core were not taken 
into account. Conversely, if the radius of the core were already known - from 
the measured tides \citep{Genova:2016aa,Konopliv:2016aa,Rivoldini:2011fk}, 
seismic sounding by the SEIS instrument on InSight \citep{Panning:2016aa}, or 
from nutation amplitudes measured by RISE and LaRa - and if the effect on the 
period of the FCN is larger than the expected precision of the LaRa 
measurements, then the dynamic flattening of the core could be constrained. 
Unlike the FCN period, $F$ does not depend directly on the dynamic flattening of 
the core and therefore core radius estimates based on $F$ are more robust than 
those based on the FCN period.
\\
Partial melt layers in the mantle due to thermal effects or water could 
significantly affect the mantle rheology and the compliances, but they affect 
the FCN, $F$, and $T_F$ only weakly since these parameters depend mainly on the 
moments of inertia of the planet and core. The effect of the rheology on 
nutations can be further reduced if prior knowledge about the mantle rheology 
deduced from the tidal dissipation factor of Mars is included in the modeling. 
Such a rheology reduces the FCN period by up to 1.5 days and $F$ by up to 3\%. 
\\

\subsubsection{Constraints on the dynamic of Mars atmosphere }

    Today, the lack of continuous global observations of dynamical variables, 
i.e.~with a good temporal and spatial resolution, limits studies of the dynamics 
of the Martian atmosphere. Although, currently,  the overall agreement between 
the observed and modeled LOD variations is fairly good \citep{Karatekin:2017aa}, 
the present knowledge of LOD is not sufficient enough to differ between models 
and to constrain Mars  CO$_2$ cycle or winds.  In addition, the accuracy of the 
observations is not sufficient to determine the polar motion or the interannual 
variability of the LOD.  LaRa will improve the current precision on LOD  more 
than an order of magnitude with respect to previous orbiter determinations\red{. 
This improved precision will} provide constraints on the details of the physical 
processes taking part in the angular momentum variations of the Martian 
atmosphere. In addition, with such a precision on LOD and with longer tracking 
coverage it might be possible to see the inter-annual variations of the 
global-scale cycling of  CO$_2$ on Mars that could arise from the dust storm 
variability. Furthermore determination of Chandler Wobble would provide 
information on the interior based on the value of the frequency as well as 
    novel information on Mars' atmospheric dynamics  in a band close to the CW 
frequency.

\subsection{Surface platform positioning}
\label{sectionposi}
\begin{figure}[!ht]
\centering
\includegraphics[width=0.9\textwidth]{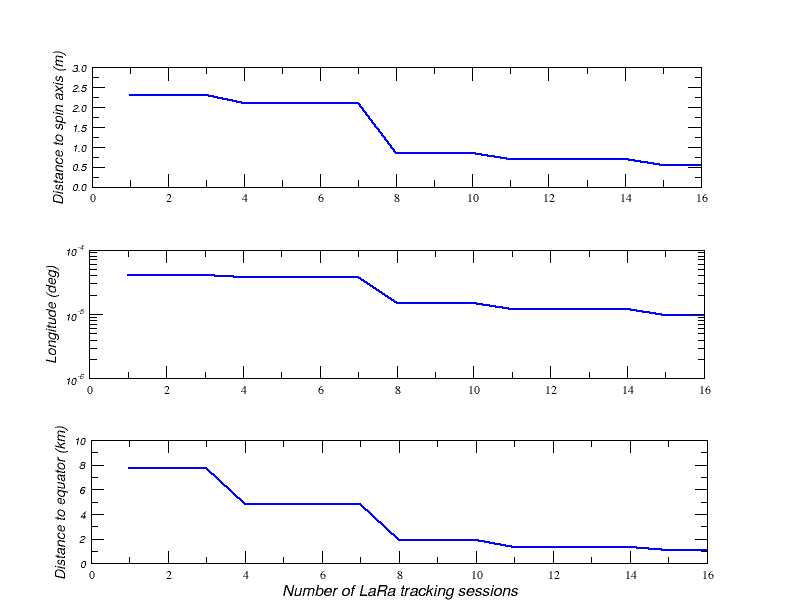}
\caption{LaRa's positioning of the ExoMars 2020 Surface Platform at the surface 
of Mars after few hours of tracking. The figure shows the time evolution of the 
uncertainties in $r_0$ (upper panel), in longitude, $\lambda_0$ (middle panel) 
and in $Z_0$ (lower panel). }
\label{fig:sigpos}
\end{figure}
\begin{figure}[!ht]
\centering
\includegraphics[width=0.9\textwidth]{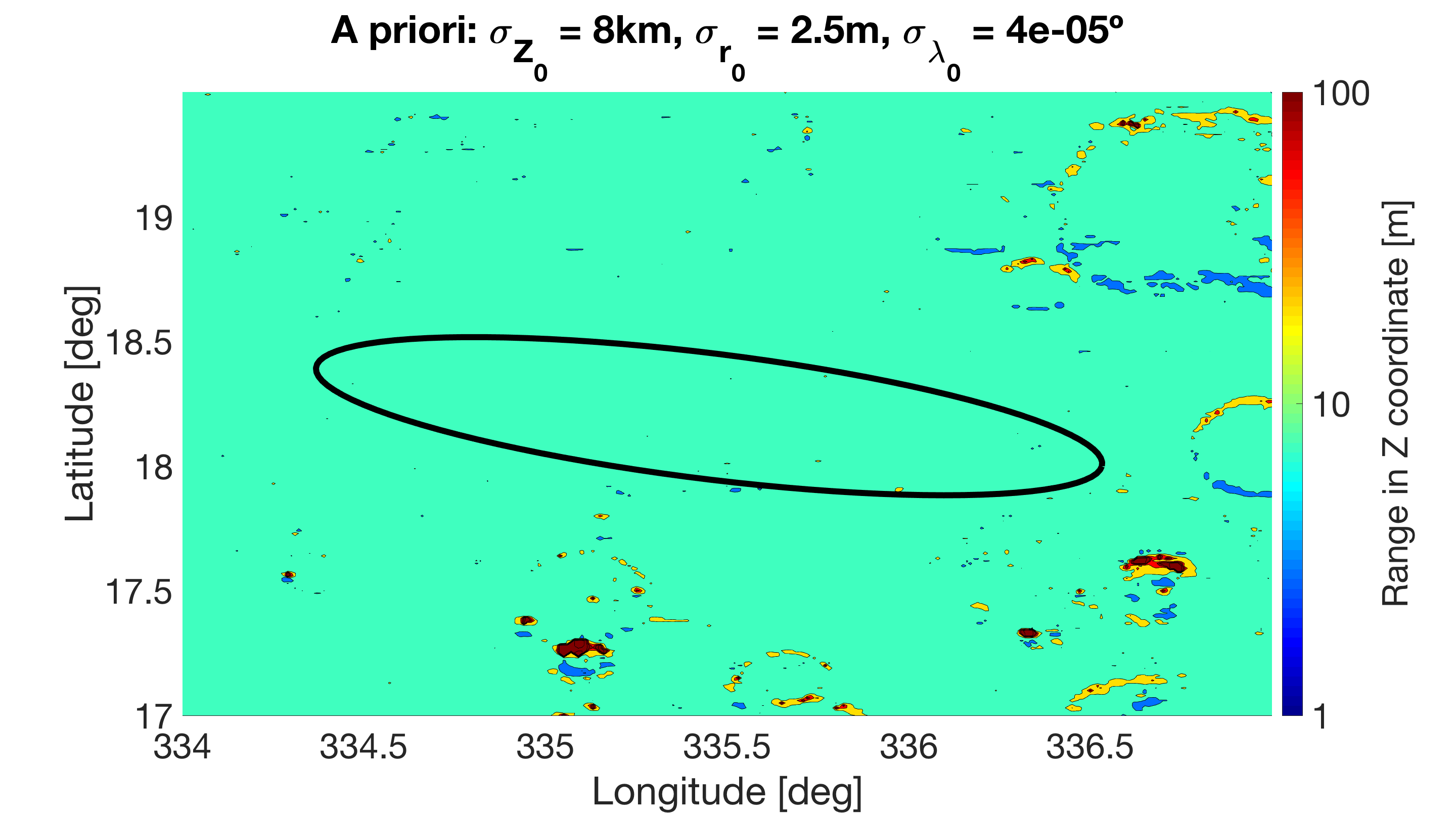}
\caption{Range of allowed Z-coordinates as a function of the actual lander 
location inferred from a priori knowledge that we get after a couple of hours of 
DTE Doppler tracking of LaRa.}
\label{fig:Zrange}
\end{figure}

Beside measuring the rotational motion of Mars, LaRa will accurately determine 
the location of the ExoMars \red{Kazachok Platform} at the surface of Mars early 
after landing. Indeed, after a few of hours of tracking, LaRa will allow 
determining the SP equatorial-plane coordinates (longitude, $\lambda_0$ and 
distance to spin axis, $r_0$) with high level of accuracy, while the $Z_0$ 
coordinate (along the spin axis) will only be estimated with 1-8 km depending on 
the actual number of tracking hours used to determine the lander location (see 
Fig.~\ref{fig:sigpos}). The inefficiency of the Mars lander's DTE Doppler 
measurements to accurately determine the $Z$ coordinate is well known 
\citep{Le-Maistre:2012ff} and can be counterbalanced by forcing the lander's $Z$ 
coordinate to tie the lander to the surface, as defined by a topography model. 
Using such a method proposed by \cite{Le-Maistre:2016aa} and successfully 
applied by the ROB team for InSight, $\sigma_{Z_0}$ will be reduced to about 10m 
(see Fig.~\ref{fig:Zrange}), which is more than two orders of magnitude smaller 
than the uncertainty provided by DTE Doppler measurements alone. \\
\\
An early estimate of the lander location will certainly improve the quality of 
the science analysis (for instance by making easier the recognition of the 
geological features seen on the lander's pictures) and allow high-resolution 
cameras from spacecraft to correctly target the lander. The advantage of 
\cite{Le-Maistre:2016aa}'s approach is that it does not rely on measurements 
from orbiters, which might take extra time to acquire and deliver to Earth the 
lander imaging data that would allow to locate the spacecraft on the Martian 
terrain of the landing site.\\
\\
After one Martian year of operation, LaRa will allow positioning the ExoMars 
2020 \red{Kazachok} Surface Platform with an accuracy of less than 10cm for the 
equatorial-plane coordinates and few meters for the $Z$ component.

\section{Discussion and conclusions}

Over the last decade, several missions involving either spacecraft orbiting Mars 
or landers or rovers on the surface have brought new understanding of the planet 
Mars, in particular on Mars’ evolution and habitability. However, the interior 
of Mars is still not yet well known. As the interior of a planet retains 
signatures of its origin and subsequent evolution, missions addressing that 
subject will allow us to answer some of the most debated questions of the 
moment. 

The Earth’s interior is mainly constrained by seismic data from an extended 
network of seismometers. However, for Mars, there is no such network, though the 
seismic activity of Mars is currently monitored by the SEIS experiment on 
InSight \citep{SEIS:2019fk}. Till now, the existence of a liquid core inside 
Mars has been deduced from measuring tides \citep{Yoder:2003fk}. In the future, 
independent determinations of the state of the core will be provided from 
measuring Mars' rotation by the RISE experiment on InSight 
\citep{Folkner:2018aa} and by the LaRa experiment on ExoMars 2020. Besides 
determining the state of the core, precise measurements of Mars' rotation and 
polar motion also allow to provide constraints about its shape and the energy in 
the atmosphere at seasonal timescale.

The paper has further addressed the understanding of the core, based on the 
determination of Mars' rotation and orientation variations. The related 
scientific experiments on fixed landers on the surface observed from Earth with 
the help of radio signals are either just in place (RISE on InSight, landed on 
November 26, 2018) or will be in place (LaRa on ExoMars 2020, to land in 2021). 
The present paper has addressed the LaRa experiment in this perspective.

In Section~\ref{sectionscirationale}, we have described the rotation of Mars, 
the models that are presently available and on which the LaRa experiment 
simulations have been based. In Section~\ref{sectionLaRaexp}, we have described 
the LaRa instrument (a coherent transponder and three antennas working in 
X-band) and presented the possibilities that it allows in terms of science. In 
particular, the MOP (Mars Orientation and rotation Parameters) signatures have 
been studied and the interior parameters that can be reached have been presented 
in Section~\ref{sectionMOP}. We have discussed these parameters and have shown 
that the core amplification factor increases with core radius - or equivalently 
with the core moment of inertia (Eq.~(\ref{eq:TF})) - and that it provides the 
best measurement of the core radius. Indeed, it is only weakly dependent on the 
thermal state of the planet and on the mantle mineralogy. With an expected 
precision of LaRa concerning the nutation amplitudes of 4~mas, the core radius 
can be estimated with a precision better than 160~km. The possibility of 
observing the FCN period from LaRa data depends on the actual value of the FCN, 
which depends on the core radius, the mantle rheology, the equilibrium state of 
the core (hydrostatic or not). The non-hydrostatic contribution was not 
discussed in this paper. It will shift the resonances to the CW and the FCN by 
about 20 days (a decrease from 220~to~200~days for the CW period and from 
$-$240~to~$-$260~days for instance for the FCN period), but will not change the 
results of this paper. This paper represents an interesting information to 
envisage the future LaRa experiment in complement with the RISE experiment.

\red{Further discussions and details on the impact of the operational and 
technical characteristics of LaRa on the nutation determination 
\citep{Le-Maistre:2019ab} and on the complementarities between the RISE 
(Rotation and Interior Structure Experiment) experiment on InSight and the LaRa 
experiment \citep{Peters:2019aa} are provided in two companion papers.}

\section{Acknowledgments}

For the Belgian authors, this work was financially supported by the Belgian 
PRODEX program managed by the European Space Agency in collaboration with the 
Belgian Federal Science Policy Office.
\\
This project was made possible thanks to a dedicated team at the 
AntwerpSpace company, including Jean-François Boone, Lieven Thomassen, Donald 
Heyman, Felix Rautschke, Dave Bail, Bart Desoete, Wim Philibert, Vincent Bath, 
Koen Puimege, and their collaborators.\\
Finally, the LaRa science team would like to warmly thank Marline Claessens, 
Daniel Firre, Jean-Philippe Halain, Alberto Busso, Marco Sabbadini, Lorenzo 
Marchetti, Michel Lazerges and ESA personnel  who supported the LaRa technical 
staff in the development of the instrument all over the project.\\
This research was carried out in part by the InSight Project at the Jet 
Propulsion Laboratory, California Institute of Technology, under contract with 
the National Aeronautics and Space Administration.

\appendix

\bibliographystyle{plainnat}

\bibliography{bibref}

\end{document}